\documentclass[twocolumn]{aastex63}
\usepackage{verbatim}
\usepackage{amsmath}
\usepackage{amssymb}

\renewcommand{\L}{\mathcal{L}}
\usepackage{bm}
\usepackage[normalem]{ulem}
\usepackage{xifthen}
\newcommand{\new}[2][]{#2}
\newcommand{\neww}[2][]{#2}

\shorttitle{Confirming ALMA Calibration using Planck and ACT Observations}
\shortauthors{Farren et al.}

\begin{document}

\title{
Confirming the Calibration of ALMA Using Planck Observations}

\author{Gerrit S. Farren}
\affiliation{Department of Applied Mathematics and Theoretical Physics\\
	University of Cambridge \\ 
	Cambridge CB3 0WA, UK}
\affiliation{Department of Physics and Astronomy\\
Haverford College \\
370 Lancaster Ave \\
Haverford, PA 19041, USA}

\author{Bruce Partridge}
\affiliation{Department of Physics and Astronomy\\
	Haverford College \\
	370 Lancaster Ave \\
	Haverford, PA 19041, USA}
	
\author{Rüdiger Kneissl}
\affiliation{Atacama Large Millimetre/submillimetre Array\\ALMA Santiago Central Offices\\Alonso de Cordova 3107\\Vitacura, Casilla, 7630355, Santiago, Chile}
\affiliation{European Southern Observatory\\ESO Vitacura\\
	Alonso de Cordova 3107\\
	Vitacura, Casilla, 19001, Santiago, Chile}

\author{Simone Aiola}
\affiliation{Joseph Henry Laboratories of Physics\\
	Jadwin Hall, Princeton University\\
	Princeton, NJ 08544, USA}
\affiliation{Center for Computational Astrophysics\\
	Flatiron Institute\\
	162 5th Avenue\\
	New York, NY 10010, USA}

\author{Rahul Datta}
\affiliation{Department of Physics and Astronomy\\
Bloomberg Center for Physics and Astronomy, Johns Hopkins University\\
3400 N. Charles Street\\
Baltimore, MD 21218, USA}

\author{Megan Gralla}
\affiliation{Department of Astronomy/Steward Observatory\\
	University of Arizona\\
	933 N Cherry Ave\\
	Tucson, AZ 85721, USA}

\author{Yaqiong Li}
\affiliation{Kavli Institute and Department of Physics\\
	Cornell University\\
	245 East Ave\\
	Ithaca, NY 14850, USA}
	
\correspondingauthor{Gerrit S. Farren}
\email{gsf29@cam.ac.uk}



\begin{abstract} 
We test the accuracy of ALMA flux density calibration by comparing ALMA flux density measurements of extragalactic sources to measurements made by the Planck mission; Planck is absolutely calibrated to sub-percent precision using the dipole signal induced by the satellite's orbit around the solar system barycenter.  Planck observations ended before ALMA began systematic observations, however, and many of the sources are variable, so we employ measurements by the Atacama Cosmology Telescope (ACT) to bridge the two epochs.  \added{We compare ACT observations at 93 and $\sim$145 GHz to Planck measurements at 100 and 143 GHz and to ALMA measurements made at 91.5 and 103.5 GHz in Band 3.  For both comparisons, flux density measurements were corrected to account for the small differences in frequency using the best available spectral index for each source.  We find the ALMA flux density scale (based on observations of Uranus) is consistent with Planck. All methods used to make the comparison are consistent with ALMA flux densities in Band 3 averaging 0.99  times those measured by Planck. One specific test gives ALMA/Planck = $0.996 \pm 0.024.$ We also test the absolute calibration of both ACT at 93 and $\sim$145 GHz and the South Pole Telescope (SPT) at 97.43, 152.9 and 215.8 GHz, again with reference to Planck measurements at 100, 143 and 217 GHz,  as well as the internal consistency of  measurements of compact sources made by all three instruments.}

\end{abstract}



\section{Introduction} \label{sec:intro}
Accurate and consistent calibration of flux density scales is obviously important to radio and sub-millimeter astronomy, especially when observations made at different frequencies, or by different instruments, are to be compared.  Ideally, such calibration would be absolute as well as accurate.  As shown by \citet{partridge_absolute_2016}, refined and absolute calibration of ground-based radio telescopes can be achieved by transferring the absolute calibration of cosmic microwave background (CMB) experiments to these other instruments.

The Planck satellite CMB mission \citep{planck_collaboration_planck_2016}, like the earlier WMAP (Wilkinson Microwave Anisotropy Probe) mission, is absolutely calibrated using measurements of the small (amplitude $\sim$0.3 mK) dipole signal induced in the $3$K CMB by the annual motion of the satellite around the solar system barycenter (see section \ref{subsec:planck_calibration} for details, or \cite{bennett_nine-year_2013} for WMAP).  Planck also has adequate sensitivity to detect thousands of Galactic and extragalactic compact sources
\citep[PCCS2;][]{planck_collaboration_planck_2016-3}.  The Planck flux densities of these sources can then be compared to flux density measurements made at ground-based instruments to check the calibration of the latter.  Such a comparison was carried out by \citet{partridge_absolute_2016} for the Karl Jansky Very Large Array (VLA) and the Australia Telescope Compact Array (ATCA).  The accuracy achieved at 22, 28 and 43 GHz was 1-2\%.  While the flux density scales of the three instruments agreed within errors at 22 and 28 GHz, a $6.2 \pm 1.4$  \% discrepancy between Planck and the VLA was found at 43 GHz (see section \ref{sec:discussion} for an update).  In this paper, we extend such comparisons to higher frequencies, with the specific aim of transferring absolute calibration from Planck to ALMA.  Note that this calibration method differs from comparing maps or angular power spectra of the CMB made by absolutely calibrated space experiments to those made by ground-based instruments such as the South Pole Telescope \citep[SPT; see][]{story_measurement_2013, crites_measurements_2015} or the Atacama Cosmology Telescope \citep[ACT; see][]{hajian_atacama_2011,louis_atacama_2014,choi_atacama_2020}.   CMB fluctuations are typically much larger in angular scale than the instrumental beams of the ground-based telescopes; in contrast, compact sources are generally smaller and unresolved.  Any comparison of the flux density scales between two instruments based on observations of compact sources hence requires (and tests) precise knowledge of the beam solid angles of both instruments.  Using and comparing direct measurements of the flux densities of compact sources also differs from calibration based on observations of the emission from planets \citep[e.g.][]{hajian_atacama_2011,butler_flux_2012}, since these depend on models of planetary atmospheres. The claimed accuracy \citep{moreno_planetary_2010} of planet-based models used for calibration of some CMB experiments (and ALMA) is $\pm 5$\%.

For three reasons, however, the comparison between Planck and ALMA measurements of compact sources cannot be made directly.  First, the central frequencies of Planck’s bands do not \replaced{necessarily}{exactly} match frequencies used at ALMA, so some interpolation or extrapolation is required.  Second, the majority of Planck compact sources observed at frequencies of 217 GHz and below are extragalactic radio sources, many of them blazars.  These are known to be variable  \citep[e.g.][]{planck_collaboration_planck_2011-1,planck_collaboration_planck_2016-3}.  Variability would not present a problem if observations of a given source could be made simultaneously by both Planck and ALMA. Planck’s High Frequency instrument (HFI; operating at 100 to 857 GHz), however, ceased operations before ALMA was commissioned.  We work around that problem here by using observations made by ACT, which was in operation during the Planck mission, and continues to observe to the present, as a bridge to connect Planck and ALMA measurements made at different times.  We first compare ACT measurements to Planck’s, then use more recent ACT measurements to check ALMA’s flux density scales. Since ACT is used solely as a bridge, any small systematic bias in the ACT measurements (say in calibration or beam solid angle) will largely cancel out.  \added{This follows since such multiplicative biases affect all sources equally, and ACT measurements are compared to both Planck and ALMA measurements.}  Since the ACT, Planck and ALMA observation are not exactly simultaneous, variability remains the dominant source of uncertainty in our results (see discussion in sections \ref{sec:internal_consistency_act_planck}, and \ref{sec:planck_based_calbration_ALMA}).  Finally, Planck, ACT and ALMA have very different resolution; that is, beams with very different solid angles.  For true point sources, that would make no difference (except that Planck’s large beams might encompass background objects other than the target source).  We address this issue in part by excluding sources for which there is evidence of resolved or confused structure; see further discussion in sections \ref{subsec:flux_comp_general_features} and \ref{subsec:linking_planck_alma}.

Section \ref{sec:calibraton_standards} treats the absolute calibration of Planck and reports the small uncertainties in calibration of the frequency bands employed here, 70, 100, 143, 217 and 353 GHz; it also  provides values for the solid angles of Planck beams.  Section \ref{sec:calibraton_standards} also describes the present standards used for calibrating ACT and ALMA.  Section \ref{sec:comp_surce_flux_densities} deals specifically with flux density measurements of compact sources made by the instruments we consider here, and with the method we adopted to interpolate or extrapolate flux densities to different frequencies. \added{In section \ref{subsec:flux_comp_general_features} we describe general features of all our flux comparisons including some methods for mitigating the effect of variable sources.} Section \ref{sec:internal_consistency_act_planck} (with details given in Appendices \ref{appendix:f2f_plack_consistency} and \ref{appendix:act_updates_note}) treats the internal consistency of both the catalogued Planck flux densities and the ACT flux density measurements that we use to link the earlier Planck results to the later ALMA results. \deleted{Some methods for mitigating the effect of variable sources are treated here as well.}  Section \ref{sec:planck_based_calbration_ALMA} compares Planck and ACT measurements; here we need to consider angular resolution for the first time, as well as the very different sensitivities of the two instruments (see also Appendix \ref{appendix:var_bias}).  Section \ref{sec:planck_based_calbration_ALMA} also presents the main result of this paper, a comparison of ALMA  and Planck flux densities of compact sources, as a direct test of the absolute flux density scale used at ALMA.  Also included are comparisons with the recently released South Pole Telescope (SPT) catalog of compact sources \citep{everett_millimeter-wave_2020}.  \replaced{We summarize \new[ and discuss some related earlier results in the final section.]{in the final section, where we also discuss some related earlier results and plans to extend this work. }}{We analyze  our various comparisons  of Planck and ALMA flux densities in section \ref{sec:discussion}, where we also discuss some related earlier work. Conclusions are briefly reported in the final section (section \ref{sec:conclusions}).  }

\section{Calibration Standards}\label{sec:calibraton_standards}

In this section, we summarize the methods that have been used to calibrate the three instruments treated in depth in this paper.  Briefly, as noted above, Planck is absolutely calibrated on the orbital dipole; ACT calibration is based on comparing its CMB maps and power spectra to Planck CMB results; and ALMA is calibrated using observations of planets, particularly Uranus.

\subsection{Planck calibration}\label{subsec:planck_calibration}

The calibration of the HFI instrument on Planck is extensively discussed in \citet{planck_collaboration_planck_2020} and earlier papers cited therein; see also \citet{planck_collaboration_planck_2016}. The calibration of the 100, 143, 217 and 353 GHz detectors is ultimately determined by the amplitude of the dipole signal induced in the CMB by the satellite’s annual orbital motion around the solar system barycenter.  The amplitude of the induced dipole, to first order $\Delta T = (v/c) T_0$, depends only on the satellite’s velocity (known to better than 0.01\% precision) and the temperature of the CMB, which we take to be  $T_0 = 2.72548 \pm 0.00057$ K \citep{fixsen_temperature_2009}.  In that sense, Planck calibration is absolute.  The small uncertainty in the calibration, as discussed in \citet{planck_collaboration_planck_2020} and listed in Table \ref{table:calibration_table}, is dominated by effects of non-linearity in the response of the instrument’s analog-to-digital converters.  
For the calibration at 70 GHz, also based on the annual dipole, see the corresponding paper for the Planck Low Frequency Instrument (LFI), \citet{planck_collaboration_planck_2020-1}.

It is important to keep in mind that Planck is calibrated on a large-angular-scale signal, a dipole distortion of the CMB.  To transfer that calibration to flux densities of compact sources requires \new[a precise]{detailed} knowledge of beam solid angles for the detectors in each Planck channel.  For that reason, we list the uncertainties in beam solid angle taken from \citet{planck_collaboration_planck_2020}.  Any error in the solid angle would introduce a systematic bias in flux density values (since $S \propto \Omega$); hence we will not add this uncertainty in quadrature to other observational errors.  From Table \ref{table:calibration_table} (column 3), however, we see that the Planck beam uncertainty lies below 0.15\% for all HFI frequencies, and is $\sim$0.5\% at 70 GHz \citep{planck_collaboration_planck_2020-1}.

Finally, the Planck flux densities we employ here are taken from two different Planck catalogs of compact sources.  In most cases, we take flux densities (“DETFLUX”) from the Second Planck Catalog of Compact Sources \citep[PCCS2;][]{planck_collaboration_planck_2016-3}\footnote{Note that \new[unless otherwise noted]{} we do not use any data from the lower quality “extended” catalog, PCCS2-E \citep{planck_collaboration_planck_2016-3}.} which makes use of the instrument calibration used for the 2015 Planck release.  The calibration has been very slightly updated for the 2018 release \citep{planck_collaboration_planck_2020,planck_collaboration_planck_2020-1}; the very small corrections are given in Table \ref{table:calibration_table} (column 6), and were applied to the PCCS2 flux densities.  PCCS2 flux densities are averaged over the entire 2.5 years HFI operated.  In contrast, the earlier PCCS1 catalog  \citep{planck_collaboration_planck_2014} provides flux densities averaged only over the period August 2009 to November 2010.  As explained in sections \ref{subsec:planck_internal_consistency} and \ref{subsec:linking_planck_alma} below, we employ the earlier PCCS1 catalog when we compare Planck to ACT flux densities for that limited period.  The PCCS1 flux densities were based on a preliminary calibration, and hence need somewhat larger 1-3\% (and 7.7\% at 353 GHz) corrections to match the 2015 calibration standard.  The correction factors are given in column 5 in Table \ref{table:calibration_table}; these include small adjustments to the calculated beam solid angles made between 2013 and 2015. 

\begin{deluxetable*}{lDDcDDDD}
    \colnumbers
	\tablecaption{Observation frequencies, beam solid angles, calibration precision and calibration shifts for Planck, ACT and SPT.  The center frequency for the ACT  measurements at $\sim$145 GHz varied \added{by less than 4 GHz} from array to array and season to season.  \added{Here and throughout, we use ``$\sim$145" to indicate ACT observations in the frequency range 144-148 GHz; the appropriate, exact, center frequency was employed in each comparison; see \ref{subsubsec:act_s2s_consistency} for an example.} \new[]{For both ACT and SPT, the center frequencies listed in column 2 assume a synchrotron spectral index.}  For ACT, the precision entered in column 4 includes uncertainty in the beam solid angle, \new[]{currently the dominant source of uncertainty.  \added{For ALMA, frequencies used in the dedicated observations are given, along with nominal flux density precision.  Finally, a positive calibration shift indicates that measured flux densities increased by that amount.}} 
	\label{table:calibration_table}}
	\tablewidth{0pt}
	\tablehead{
		\colhead{Instrument} & \multicolumn2c{Frequency,} & \multicolumn5c{Beam Solid Angle,} & \multicolumn2c{Precision,} &
		\multicolumn4c{Calibration  Shift}\\
		\colhead{} & \multicolumn2c{GHz} & \multicolumn5c{arcmin$^2$} & \multicolumn2c{\%} &
		\multicolumn2c{2013$\to$2015}&\multicolumn2c{2015$\to$2018}
	}
	\decimals
	\startdata
	Planck& 70.4 & $200.9$ & $\pm$ & $1.0$ & $0.5$ & $+1.05\%$ & $-0.015\%$ \\
	Planck& 100 & $106.22$ & $\pm$ & $0.14$ & $<0.1$ & $+1.79\%$ & $-0.02$\% \\
	Planck& 143 & $60.44$ & $\pm$ & $0.04$ & $<0.1$ & $+2.40\%$ & $-0.06\%$ \\
	Planck& 217 & $28.57$ & $\pm$ & $0.04$ & $0.2$ & $+2.39\%$ & $+0.10\%$ \\
	Planck& 353 & $27.69$ & $\pm$ & $0.02$ & $0.5$ & $+7.70\%$ & $+1.00\%$ \\\hline
	ACT& 93.0 & 6.2 &  & . & 2.6 & . & . \\
	ACT& $\sim$145 & 2.4 &  & . & 2.6 & . & . \\
	ACT& 218 & 1.8 &  & . & 2.6 & . & . \\\hline
	SPT& 97.43 & 3.3 & & . & 1.05 & . & . \\
	SPT& 152.9 & 1.6 &  & . & 1.15 & . & . \\        
	SPT& 215.8 & 1.1 &  & . & 2.24 & . & . \\ \hline
	ALMA& 91.5 & . &  & . & 5 & . & . \\
	ALMA& 103.5 & . &  & . & 5 & . & . \\
	ALMA& 343.5 & . &  & . & 10 & . & . \\
	\enddata
\end{deluxetable*}

\subsection{ACT calibration}\label{subsec:ACT_calibration}

The initial calibration of ACT data is based on observations of Uranus \citep{choi_atacama_2020}. The final calibration, however, is ultimately tied to the CMB by directly comparing ACT maps or angular power spectra of the CMB to those made by WMAP \citep{hajian_atacama_2011} and later by Planck \citep{louis_atacama_2014,choi_atacama_2020} over a range of angular scales.  \cite{louis_atacama_2014} discusses the process in detail, and also provides estimates of the accuracy of ACT calibration.  It also examines Planck-ACT  calibration based on some preliminary observations of compact sources at 148 and 218  GHz; we compare those results with those determined here in Section \ref{sec:discussion}.  Since the publication of the paper by  \cite{louis_atacama_2014},  the ACT flux density scales in the 90 and 150 GHz bands have been slightly altered \citep{choi_atacama_2020}.  The updated estimates of the calibration precision for each band are given in Table \ref{table:calibration_table}, as are the center frequencies appropriate for observations of synchrotron sources.  We include comparable quantities for the South Pole Telescope \citep{story_measurement_2013,everett_millimeter-wave_2020}.                       	

\subsection{ALMA calibration}\label{subsec:ALMA_calibration}

\added{Most of the ALMA measurements employed here were made in Band 3 (where frequencies between 84 and 116 GHz are available) and Band 7 (275 to 373 GHz).}  ALMA’s flux density calibration employs Uranus as the reference solar system object. Uranus was included in a set of dedicated observation made in 2016 November described below.  Based on the CASA planetary model \added{for the brightness temperature of Uranus} \citep{butler_flux_2012}\footnote{This 2012 model for Uranus is based on atmosphere modeling and on fluxes from various missions including flyby missions. It was used for Herschel-SPIRE calibration. The uncertainties include cloud formation and the slow rotation of the polar caps.  When Uranus is marginally resolved, the surface brightness structure complicates the zero spacing extrapolation.}, we obtain total flux densities for this epoch of $8.26$, $9.96$ and $70.55$ Jy for the primary ALMA frequencies employed here, $91.5$, $103.5$ and $343.5$ GHz, respectively.  \added{The absolute accuracy is given as 5\% peak-to-peak.}  

We note that Uranus is largely resolved in these observations on most baselines of the 12-meter configuration. While this can introduce a dependency on the accuracy of the structural model for Uranus, a sufficient number of short baselines \added{(8 with baselines between 20 and 50 m.)} was available for flux calibration for the present work. In addition, sources observed during the campaign of observations for this paper were also closely monitored before and after the campaign with the 7-m Array.  \replaced{ and the scale of their flux densities can thus be compared.}{These 7-m array observations barely resolved Uranus and allowed us to verify the agreement with the flux scale of the campaign by comparing observations of the same sources.}

\section{Compact Source Flux Densities}\label{sec:comp_surce_flux_densities}

In this section, we describe briefly how flux densities of compact sources are determined from observations made by the instruments treated here. 

\subsection{Source extraction and flux density measurements} \label{subsec:source_extraction_flux_density_measurement}

Planck, ACT and SPT compact sources are identified in sky maps using a spatial filter to minimize the effect of CMB and other large-scale fluctuations as well as noise.  In the case of Planck, for instance, a Mexican Hat Wavelet filter is used to reduce both large-scale structure and small-scale noise \citep{planck_collaboration_planck_2016-3}.  ACT employs a similar approach using a matched filter \citep[see][]{marsden_atacama_2014,gralla_atacama_2020}; for SPT, see \citet{mocanu_extragalactic_2013, everett_millimeter-wave_2020}.  Flux densities for significant detections are then determined.  \added{The first step is to convert the standard map units of temperature to flux density units such as MJy/ster; this step requires a knowledge of the effective center frequency.  The next step is to multiply by the beam solid angle; this step requires detailed knowledge of the instrument's beam solid angle.}   Given the broad bands employed by all three instruments, \added{typically $\sim25\%,$} both the beam solid angle and the effective frequency are slightly dependent on the source spectral index. These small biases and their uncertainties are accounted for either by small changes in the effective center frequency (in calculating it for ACT, we assume a spectral index of -0.5) or by color-correcting the flux densities; see the following subsection. In the case of Planck, several different methods are employed to estimate flux densities; we adopt DETFLUX as recommended in \cite{planck_collaboration_planck_2016-3} for compact sources at frequencies 353 GHz and below. 
The measured flux densities of low signal-to-noise sources can be biased, artificially boosted by noise spikes \citep[see][]{crawford_method_2010,gralla_accounting_2020}.  Consequently, flux densities of low S/N sources are frequently {``de\nobreakdash-boosted”} to their nominally true value; see \cite{gralla_accounting_2020}.  In this work, however, we do not employ de\nobreakdash-boosting, even though some de\nobreakdash-boosted flux densities are available for ACT. For ACT, SPT and ALMA, the sources we can use to compare to Planck measurements are all bright and high S/N (typically $>$10), so de\nobreakdash-boosting is not required.  The Planck measurements, on the other hand, have lower S/N,  but since de\nobreakdash-boosted values are not available, we address this potential bias in a different way (see section \ref{subsec:flux_comp_general_features}).

\subsection{Color corrections}\label{subsec:color_correction}

The instruments treated here have substantial bandwidths, \added{ approximately 25\%}. Derived flux densities can thus depend on the source spectral indices $\alpha$ (we use the convention $S\propto \nu ^\alpha$).

In the case of Planck HFI, for instance, the catalogued flux densities implicitly assume that all sources have  $\alpha = -1$, and that the effective band centers are exactly 100, 143, 217 and 353 GHz \citep{planck_collaboration_planck_2014}. Small (a few percent or less) color corrections are needed, since virtually all of the sources we consider have flatter \new[]{(closer to zero)} synchrotron spectra instead.  This issue is treated further in section \ref{subsec:planck_flux_densities} below.  At ALMA, bandpass correction is an integral part of the calibration procedure.  For the other two ground-based instruments, the required, spectrum-dependent, ``color corrections” are smaller and are made by an appropriate adjustment of the effective center frequency in each band, as well as small, additional corrections depending on the source spectrum \citep[for ACT, see][]{datta_atacama_2019,choi_atacama_2020}. The center frequencies for ACT and SPT listed in Table \ref{table:calibration_table} are those appropriate for compact sources with synchrotron spectra, \added{with spectral indices -0.7 or -0.5.}

\subsection{Planck flux densities}\label{subsec:planck_flux_densities}

All of the raw (uncorrected) Planck flux density measurements considered here come either from the PCCS1 \citep{planck_collaboration_planck_2014}, which includes measurements made from August 2009 to November 2010, or the PCCS2 \citep{planck_collaboration_planck_2016-3}, which includes measurements of flux density averaged over the entire mission (August 2009 to January 2012 for the HFI instrument).  In the case of the PCCS2, we employ only sources that appear in the main, high reliability catalog (see \citet{planck_collaboration_planck_2016-3} for details on the criteria employed as well as source extraction).  Flux densities in these catalogs are averages over all observations of a given source.  Some sources are observed at a given frequency for a few days every six months, that is once every survey.  Others, near the ecliptic poles, are observed much more frequently, given Planck’s scan strategy.  Since the two catalogs cover different (though partially overlapping) periods, we can use a comparison between flux densities for a given source in the two catalogs to flag some strongly variable sources. 

As noted above, cataloged Planck HFI flux densities need to be color-corrected unless the source in question has a $\nu^{-1}$ spectrum.  The color-corrections are a smooth function of the source spectral index, and are generally smaller than a few percent \citep{planck_collaboration_planck_2014}.  We use Planck data to determine a preliminary spectral index at each frequency for each source, and then use these spectral indices to color correct the raw flux densities.  These initial spectral indices for Planck’s 143 GHz data, for instance, are determined by comparing flux densities at the neighboring frequencies of 100 and 217 GHz.  \added{Note that this method assumes a power law spectrum for the sources.  While this may not be exact in every case, it is adequate for our purposes given how small the color corrections are.}  The color-corrected flux densities are then used to re-compute corrected and final spectral indices for each source at each frequency.  These corrected spectral indices are used when we make extrapolations or interpolations from Planck’s center frequencies of 70.4, 100, 143, 217 and 353 GHz to the frequencies used in ground-based observations. While we list both raw and color-corrected flux densities for Planck in Table \ref{table:PCCS1_v_MBAC_comp}, we plot only the latter, and use only the color-corrected flux densities to compare with results from other instruments.

\subsection{ACT flux densities}\label{subsec:ACT_flux_densities}

The ACT maps treated here are calibrated in two steps. An initial calibration in temperature units is based on observations of Uranus, array by array and season by season. These preliminary calibrations are then adjusted by a small amount so that the ACT power spectra match Planck’s CMB power spectrum in the overlapping range of spatial frequency \cite[see ][and references therein]{aiola_atacama_2020,choi_atacama_2020}.  In all cases (except for 144.1 GHz measurements by one array in 2016; see Section \ref{subsec:act_internal_consistence}), these adjustments are at the level of a few percent, and are measured to ${\sim}1$\% accuracy.  Thus  the calibration of ACT maps is ultimately tied to the absolute calibration of Planck.  A matched filter is then applied to these calibrated maps, and the amplitude of each compact source in the filtered map is measured.  These values are then converted to flux density in Jy, taking account of the beam solid angle and the frequency response of each array in each season, \added{as outlined in \ref{subsec:source_extraction_flux_density_measurement} above.}  As discussed in greater detail in \cite{gralla_measurement_2014}, multiple, band-dependent sources of uncertainty contribute to the overall uncertainty in the calibration of the ACT point source flux densities. These include uncertainty in the beam solid angle, uncertainty introduced by the mapmaking procedure, and uncertainty in the initial temperature calibration obtained from comparison to Planck \citep{choi_atacama_2020}. Combining these yields the overall uncertainties cited in Table \ref{table:calibration_table}.


\subsection{ALMA flux densities}\label{subsec:ALMA_flux_densities}

The flux densities measured by ALMA and considered here derive from three different  sources.   ALMA, as part of routine observatory calibration, has a ``grid monitoring” program of approximately two week cadence for the flux density of 40 bright quasars, which are used in PI science observations as secondary flux calibrators \citep{fomalont_calibration_2014,remijan_alma_2019}. When these are present in the ACT search area, monitoring measurements of these sources can be compared to ACT measurements made \added{in the same year}; we will refer to such measurements as “grid” observations.  

In addition, some weaker quasars and other radio sources are monitored irregularly in the vicinity of science targets depending on the scheduling of individual PI projects in order to check their suitability as phase calibrators (``cone-searches”), especially for high frequency and long baseline observations. These are more heterogeneous in both frequency and quality.  They span the interval from 2012 to 2017, and are thus subject to source variability  \citep{guzman_stochastic_2019}.  For these reasons, we treat them separately when we compare ALMA results to those from ACT (section \ref{subsec:including_other_alma_obs}).  We use only late 2015 and early 2016 ALMA observations of this kind when comparing to 2016 ACT data.  In addition, we accepted only those sets of measurements that included at least two ALMA frequency bands, so that we could establish meaningful spectral indices from these data.  When used, we refer to these lower-quality measurements as “secondary” observations. 

Finally, we rely mainly on measurements of a set of bright, extragalactic sources made in a dedicated program of ALMA observations in Bands 3 and 7  as part of the calibrator flux density update campaign\new[ and coordinated by one of us (RK)]{}; these we call “dedicated” observations.  \new[To facilitate a direct comparison between ACT and ALMA,]{ Specifically,} we observed selected compact sources that lay in the ACT \added{survey area} in a special, cone-search-like, calibrator run on $\sim$50 ALMA phase calibrators.  These targeted observations were carried out between 8 and 10 November, 2016, during ACT’s 2016 observing season, and resulted in 37 sources available for comparison to ACT. These dedicated observations, centered at 91.5, 103.5 and 343.5  GHz, used mainly the 12-m Array in configuration C40-6, with the 64-input correlator, and included Uranus as the reference calibration object.

Since we had simultaneous observations at widely different frequencies, we employed spectral indices derived from these dedicated ALMA observations\footnote{Specifically, for each source, we calculated spectral indices from 91.5 to 343.5 GHz, and from 103.5 to 343.5 GHz, then took the inverse-variance weighted average of the two. \added{We verified that the spectral indices computed for different frequency pairs are consistent within error and find excellent agreement at the fraction of a $\sigma$ level.}} to interpolate (or extrapolate) to ACT’s central frequencies of 93 and $\sim$145 GHz, and to Planck’s central frequencies of 100, 143, 217 and 353 GHz.  We tested the results by employing spectral indices derived instead from ACT 93 and $\sim$145 GHz observations (see Section \ref{subsubsec:act_alma_comp_w_act_alpha}).  Typical uncertainties in these dedicated ALMA measurements were 10-20 mJy, and up to $\sim$100 mJy for the brightest sources, i.e. very roughly 1/2 the Planck uncertainties.  As expected, flux density uncertainties for the ''secondary” observations were often larger.  See section \ref{subsec:linking_planck_alma} for further details.  \added{The uncertainties we cite include the standard 5\% calibration uncertainty (10\% at 343.5 GHz) assumed for ALMA observations.  For the bright sources employed here, this multiplicative source of error dominated.  It is this calibration uncertainty we seek to reduce.
}

The absolute calibration of all three data sets is based on Uranus, as noted in section \ref{subsec:ALMA_calibration} above.  We note that the brightness temperature of Uranus has been confirmed by absolutely-calibrated measurements made by WMAP \citep{weiland_seven-year_2011}.

\section{General features of all flux density comparisons}\label{subsec:flux_comp_general_features}

\added{The main goal of this paper is to compare flux densities measured by one instrument (or at one time) with flux densities measured by another instrument (or at a different time). We thus begin in here with a brief description of how we made these flux/flux comparisons.} As noted, the instruments considered here operate at many different frequencies.  We therefore need to interpolate or extrapolate flux densities to obtain a proper match.  To do so, we use the best available spectral index for each source, as described in detail for each comparison. In most cases, these are based on the multi-frequency Planck results (see section \ref{subsec:color_correction}); in some cases, we employ ACT-based or ALMA-based values (section \ref{subsec:ALMA_flux_densities}).  We also record for each comparison the frequency at which the comparison is made.  Except for ALMA data in some bands, these extrapolations involve adjustments to the flux density of a few percent or less.

The instruments also have different detection thresholds: Planck in particular is much less sensitive to compact sources than the larger ground-based instruments.  As noted in section \ref{subsec:source_extraction_flux_density_measurement}, this raises the possibility of Eddington bias or other biases when we compare Planck results with ground-based results.  As in \citet{partridge_absolute_2016}, we mitigate the effect of bias by forcing the fit in flux density vs flux density plots to pass through (0, 0).  Unless otherwise specified all the figures and results cited below are subject to this constraint.  Thus, the slope of the resulting linear fit provides a direct comparison of the flux density scales of the two data sets being compared; a slope of unity is expected for flux density scales in perfect agreement.

\subsection{Accommodating variability}\label{subsubsec:accommodating_variability}

As noted in section \ref{sec:intro} above, most of the Planck sources are AGN, and therefore likely to be variable.  Since flux densities can either increase or decrease, variability should not on average introduce a bias in our comparisons, but it will increase the scatter.  The scatter in most of the comparisons shown in figures in this paper is dominated by source variability. We attempted to reduce the effect of variability-induced scatter on flux/flux plots by first fitting a line to all the data, then dropping outliers and re-computing the slope of the fit.  Specifically, we first fit all the data using orthogonal distance regression (ODR, \added{see \cite{boggs1990orthogonal}}). This model assumes that given any observational data point $(\hat{x}_i, \hat{y}_i)$, where both $\hat{x}_i$ and $\hat{y}_i$ are subject to error, there exist true values $(x_i, y_i)$ such that $y_i=f(x_i;\beta)$. $f$ is some fitting function dependent on parameters $\beta$ (In our case the fitting function is simply $f(x; m) = m x$). The observed values are related to the ``true" values as $x_i =\hat{x}_i + \delta_i$ and $y_i = \hat{y}_i + \epsilon_i$ where $\delta_i$ and $\epsilon_i$ are unknown. 

The best fit line is determined by minimizing the loss function
\begin{equation}
	\L(\bm{\delta},  \bm{\epsilon}) = \sum_i \delta_i^2 + \epsilon_i^2
\end{equation}
over all $\delta_i$ and $\epsilon_i$.
Since we assume that $y_i=f(x_i;\beta)$ we can express this in terms of   our observations, the fitting function and the parameters $\beta$ as
\begin{equation}
	\L(\bm{\delta},  \beta) = \sum_i \delta_i^2 + \left[f(\hat{x}_i+\delta_i;\beta) - \hat{y}_i\right]^2.
\end{equation}
In order to account for unequal errors in the two data sets and between different data points we inverse-variance weight this loss function
\begin{equation}
	\L_w(\bm{\delta},  \beta) = \sum_i \frac{\delta_i^2}{\sigma_{\hat{x}_i}^2} + \frac{1}{\sigma_{\hat{y}_i}^2}\left[f(\hat{x}_i+\delta_i;\beta) - \hat{y}_i\right]^2.
\end{equation}
The fitting is performed using the ODR routines available in \textit{SciPy} \citep{virtanen_scipy_2020} which are based on the Fortran package ODRPACK \citep{boggs_algorithm_1989}.

To perform the iterative fitting we compute the weighted root-mean-square (rms) distance of all observed data points from their respective ``true" values
\begin{equation}
	d_{\text{rms}}= \sqrt{\frac{1}{N}\sum_i \left(\frac{\delta_i^2}{\sigma_{\hat{x}_i}^2} + \frac{\epsilon_i^2}{\sigma_{\hat{y}_i}^2}\right)}.
\end{equation}
Here $N$ is  the total number of data points fit in each iteration. We then iteratively exclude all data points with a weighted distance $d_i=\sqrt{\frac{\delta_i^2}{\sigma_{\hat{x}_i}^2} + \frac{\epsilon_i^2}{\sigma_{\hat{y}_i}^2}}$ of more than three times $d_\text{rms}$ until convergence is reached.  In the figures, the sources dropped in this iterative process (“outliers”) are shown in grey. In most cases only a small number of iterations (0-5) were required before the slope of the fit converged to a final value.  In all cases, the number of sources dropped by this iterative process, and thus not included in the final fits, was a small fraction of the total.

\subsection{Sample variance}\label{subsubsec:sample_variance}

\added{To estimate the effect of sample variance on our flux/flux fits, we employed a bootstrapping method to the comparisons included in the chain linking Planck to ACT and thus to ALMA. Given a set of sources for comparison, we drew 2000 random samples from the original set, assigning equal weight to all data points. To retain the iterative method described above, we first performed an iterative fit on the full data set and then re-sampled only those data points which remain included in the fit after the iterative procedure terminated. We thus excluded the small number of outliers.  Consequently, we did not need to iterate the fit for any of the bootstrap samples. We manually inspected convergence of the mean slope and variance of these bootstrap samples and in all cases found the mean to be in excellent agreement with the values coming from the procedure described above in \ref{subsubsec:accommodating_variability}. In almost all cases, the distribution of the 2000 bootstrap slopes was close to Gaussian. Not unexpectedly, the variance found via this approach ran somewhat higher than the statistical estimate from the ODR routine, particularly for comparisons with small sample size. We employed this bootstrap approach to all flux/flux comparisons with a small number of sources (under 300).  We elected not to include this procedure in any of the internal Planck or ACT comparisons since the number of sources is large in all those cases and hence no significant contribution from sample variance is expected. It should also be noted that we assumed that the sample variances for different comparisons in our chain are independent. This assumption is overly conservative, leading to slightly inflated errors, since we expect the partial overlap between the samples used in different comparisons to introduce some 
level of  non-zero covariance. Where we do include sample variance we report the \textit{additional} contribution to the uncertainty on the calibration as $m\pm \sigma_{\rm{stat.}} \pm \sigma_{\rm{sample}}$\footnote{$\sigma_{\rm{sample}}$ is the difference between the bootstrap estimate of the error and the statistical estimate from the ODR routine.} and propagate it appropriately.}

\subsection{Dropping known variable and resolved sources}\label{subsubsec:dropping_bad_sources}

We also excluded a small number of bright, extended or known variable sources at the outset from all flux/flux comparisons and all plots shown here.  Specifically, we excluded 33 bright (and often extended) sources such as Cen A, PKS 0521-36, the Orion Nebula, 3C279, Pictor A and 3C454.3 from all the comparisons made here. The number of sources excluded from each comparison varies, however, since not all of these sources are located in sky regions accessible to ACT, SPT or ALMA.

\section{The Internal Consistency of ACT and Planck Flux Density Measurements}\label{sec:internal_consistency_act_planck}

\replaced{The main goal of this paper is to compare flux densities measured by one instrument (or at one time) with flux densities measured by another instrument (or at a different time). We thus begin in section \ref{subsec:flux_comp_general_features} with a brief description of how we made these flux/flux comparisons, then turn to the issue of internal consistency in the Planck and ACT data.}{In this section we turn to the issue of internal consistency in the Planck and ACT data.} The ACT measurements were made over several different observing seasons (a season at ACT typically spans April through the following December or early January), and employed several different receiver arrays as the instrument was made more sensitive step-by-step over the past decade \citep[see][]{swetz_overview_2011,thornton_atacama_2016,henderson_advanced_2016}.  The telescope and optics, however, remained largely the same so that the solid angle of ACT’s beam at a given frequency did not change much from season to season. Beam solid angles were assessed each season \added{by making scans of planets and other bright, unresolved sources.  The estimated uncertainty in beam solid angle is 2\%; recall that this systematic uncertainty cancels out when we compare ACT to both Planck and ALMA.}  
The small changes in beam solid angle were taken into account when computing the flux densities of compact sources.  While some of the ALMA observations used here were made at scattered times between 2012 and 2017, the underlying calibration of ALMA remained fixed, based on the model for Uranus.  Planck flux density measurements were made over the entire span of the mission, with no instrumental changes.  Nevertheless, since we will be comparing measurements made at different epochs, we need to ensure that data from each instrument are internally consistent over time, and between frequencies.

\subsection{Internal Consistency of Planck data}\label{subsec:planck_internal_consistency}

\begin{figure}
	\plotone{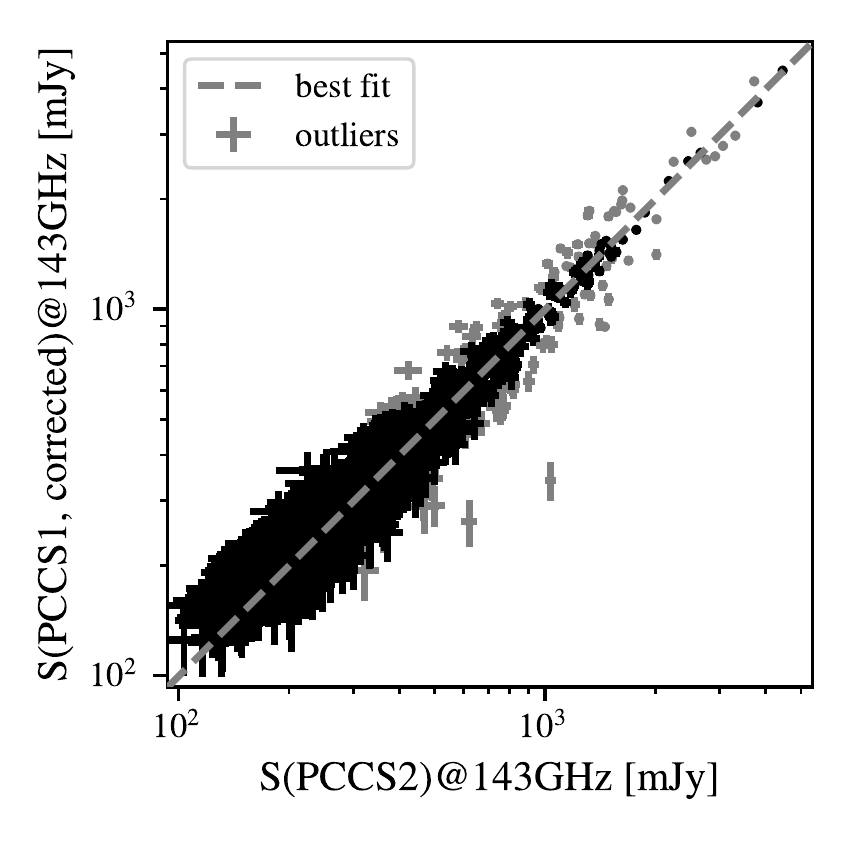}
	\caption{Corrected flux densities of 1318 sources from the 143 GHz PCCS1 (vertical axis) compared to flux densities from the PCCS2, again at 143 GHz.  Scatter is caused mostly by source variability; 95 outlier sources shown in grey are omitted from the final fit.  The slope of the resulting best-fit line (shown)
	is $0.9952 \pm 0.0026$.  Note that for clarity, we plot $\log(S)$, even though the fits were made to linear data.  \label{fig:143GHZ_PCCS1_v_PCCS2}}
\end{figure}

In the case of Planck, we do not have year-by-year or survey-by-survey flux density measurements available.  Instead, as noted in section \ref{subsec:source_extraction_flux_density_measurement}, we have flux densities averaged over the first $\sim$15 months of Planck observations, as well as flux densities averaged over the entire ($\sim$30 month) lifetime of the HFI instrument (Aug. 2009 to Jan. 2012).  These are listed in the Planck Catalog of Compact Sources \citep[PCCS1; see][]{planck_collaboration_planck_2011-1} and the Second Planck Catalog of Compact Sources  \citep[PCCS2; see][]{planck_collaboration_planck_2016-3}, respectively.  We do not expect flux densities for a given source drawn from the two catalogs to agree exactly, both because of instrument noise and because of possible variability in the sources.  We can, however, ask whether, on average, flux densities of a large sample of sources were the same in the two catalogs.  In the process of making this consistency check, we can see if the Planck measurements themselves reveal sources that varied strongly during the HFI observations; these appear as outliers in the plots of PCCS1 vs PCCS2 flux densities such as Fig. \ref{fig:143GHZ_PCCS1_v_PCCS2}.

We performed this test by matching sources in PCCS2 with sources in the older PCCS1 at each frequency, using a search radius of $6^\prime$.  From the outset, 33 bright, extended sources were excluded. The number of remaining matches ranged from $\sim$700 to $\sim$1300, depending on frequency.  

To make a fair comparison of flux densities, we take into account small changes in calibration and beam solid angle made by the Planck team between the 2013 release and the later 2015 release.  As discussed in \citet{planck_collaboration_planck_2016}, introducing the orbital dipole  as the basis for calibration and a few other small issues changed the calibration coefficients by a few percent for the Planck frequencies considered here.  At 143 GHz, for instance, the effect of these small changes was to increase flux densities by 1.00\% from the 2013 data release to the 
2015 release.  In the case of the HFI instrument, there were also small changes in the effective beam solid angle which affected the calculation of flux densities of compact sources.  At 143 GHz, the increase in the beam solid angle was 1.384\%.  The combined shift in the flux density scale at 143 GHz is thus +2.40\%; see Table \ref{table:calibration_table} for corresponding shifts for other Planck bands.  These changes in calibration were applied to the older PCCS1 flux density values before they were compared to either the PCCS2 values (as in Fig. \ref{fig:143GHZ_PCCS1_v_PCCS2}) or to ACT and ALMA measurements.  There were also much smaller shifts in calibration between the 2015 and the final 2018 release of Planck data (col. 6 of Table \ref{table:calibration_table}); we take the 2018 calibration as our standard, and have corrected both PCCS1 and PCCS2 values accordingly. 

As Fig. \ref{fig:143GHZ_PCCS1_v_PCCS2} demonstrates, the slope of the fit for 143 GHz is clearly close to unity, but there are a number of outliers (shown in grey in the figure).  We assume these are variable sources.  The slope of the fit to all the data is $1.003 \pm 0.004$; if we exclude 95 outliers,  presumably variable sources, using the $3\sigma$ criterion of section \ref{subsec:flux_comp_general_features}, the slope settles to $0.9952 \pm 0.0026$.  The final adjustment is to add the 0.07\% uncertainty in the beam solid angle discussed in section \ref{subsec:planck_calibration}.  We end with a slope of $0.9952 \pm 0.0033$. The flux density scales for the two Planck catalogs \new[are in excellent agreement]{agree} at 143 GHz;  the older PCCS1 measurements (after calibration corrections from Table \ref{table:calibration_table}) lie just  $0.48 \pm 0.33$ \% below the newer PCCS2 ones. \added{For these internal comparisons, all involving very large numbers of sources, we do not report the sample variance since it is small and subdominant compared to the uncertainty in other steps of our comparison.}

\begin{deluxetable}{DDcDc}
	\tablecaption{Flux densities for the two Planck catalogs are compared, after correction for known changes in calibration and beam solid angle.  We take PCCS2 and the 2018 Planck calibration as the standard, and hence have corrected the earlier PCCS1 data by the small factors given in Table \ref{table:calibration_table} before making the comparison. \added{The final column shows the number of sources used in the comparison. Numbers in parentheses indicate the number of sources dropped (''outliers") in the iterative process described previously.}\label{table:PCCS1_v_PCCS2_comp}}
	\tablehead{
		\multicolumn2c{Frequency, GHz} & \multicolumn5c{$S($PCCS1$)/S($PCCS2$)$} & \colhead{N}
	}
	\decimals
	\startdata
	70.4 & 1.0037 &$\pm$& 0.0059 & 749(60)\\
	100 & 1.0085 &$\pm$& 0.0044 & 1074(69) \\
	143 & 0.9952 &$\pm$& 0.0033 & 1318(95) \\
	217 & 0.9726 &$\pm$& 0.0045 & 1220(76) \\
	353 & 1.0193 &$\pm$& 0.0036 & 778(25) \\
	\enddata
\end{deluxetable}

We repeated this comparison for the four other Planck bands employed here, 70.4, 100, 217  and 353 GHz (Table \ref{table:PCCS1_v_PCCS2_comp}).  Again, flux densities from PCCS1 were corrected by the small changes in overall calibration (columns 5 and 6 of  Table \ref{table:calibration_table}) and the errors include uncertainty in the beam solid angle.  With the exception of 217 and 353 GHz measurements, we see that the two Planck catalogs of compact source flux densities are closely compatible, once small, known changes in overall calibration between 2013 and 2018 are accounted for.  We have also measured with sub percent precision the small residual differences (e.g., $0.48 \pm  0.33$ \% at 143 GHz).  These small, additional correction factors are included when we compare Planck flux densities from PCCS1 to those from ground-based instruments in section \ref{sec:planck_based_calbration_ALMA}.

\subsubsection{Frequency-to-frequency consistency of Planck measurements of compact sources}\label{subsubsec:f2f_consistency_planck}

We also confirmed the band-to-band consistency of Planck flux density measurements of compact sources.  As shown in \citet{planck_collaboration_planck_2020}, there is agreement at the 0.5\% level or better between CMB anisotropy measurements made at 100, 143 and 217 GHz.  As noted in section \ref{sec:intro}, however, the CMB signal is typically larger than the beam size at these frequencies unlike the case for compact sources; does the good inter-frequency agreement extend to Planck measurements of compact sources?  That is examined in detail in Appendix \ref{appendix:f2f_plack_consistency}, where we compare measurements from one frequency band to predictions based on interpolation from the two neighboring frequency bands.  To test the consistency of 143 GHz measurements, for instance, we use the corrected flux densities of sources at 100 and 217 GHz.  

For 143 GHz, we find a ($1.0 \pm 0.2$\%) \% difference between the measured and predicted values; Planck measurements of compact sources at 100, 143 and 217 GHz are thus consistent at a percent level.  At 217 GHz agreement is also acceptable at ($1.7 \pm 0.5$\%).  The situation at 100 GHz is less good, with the measured flux densities running on average ($2.7 \pm 0.3$\%) higher than the predicted values.  We suggest in Appendix \ref{appendix:f2f_plack_consistency} that this discrepancy could be caused by slight curvature, or a break, in the spectra of the sources used in the test. 

\subsection{ Internal Consistency of ACT data}\label{subsec:act_internal_consistence}

\new[]{As noted above, different receiver arrays were employed as ACT evolved.  In 2008-10, an early receiver array called MBAC \citep{swetz_overview_2011} was used.  It was replaced by new receiver arrays in 2012 \citep[see][]{thornton_atacama_2016}.  The ACT maps for 2013 to 2016, array by array and season by season, form part of data release DR4; see \cite{aiola_atacama_2020}.  When flux densities derived from the various season-by-season maps were averaged, we used inverse variance weights to produce the source lists such as the ``ACTall" list defined in  \ref{subsubsec:act_s2s_consistency}.  \new[At $\sim$145 GHz, the merged source list was produced by SA; at 93 GHz by YL.]{}  In contrast, the flux densities derived from the earlier MBAC runs \citep{marsden_atacama_2014,gralla_atacama_2020} are archived values drawn from LAMBDA\footnote{\url{https://lambda.gsfc.nasa.gov/product/act/act_point_sources_get.cfm}}}, the NASA Legacy Archive for Microwave Background Data Analysis.

Given these changes in the detectors employed at ACT (and some changes in the analysis pipeline as well), we investigated whether ACT flux density measurements of compact sources, once properly calibrated, were internally consistent across arrays, sky area and time.  For instance, observations during the 2013, 2014 and 2015 seasons were made with two different, polarization-sensitive, receiver arrays (called PA1 and PA2), with very slightly different operating center frequencies near 144.5 GHz. \added{A third array, PA3, including 93 GHz detectors, was added in 2015; it also operated in the $\sim$145 GHz band with a center frequency of 144.1 GHz.  In contrast, the earliest ACT measurements} (in 2008-2010) were made with an entirely different instrument, MBAC \citep[see][]{swetz_overview_2011} with an effective center frequency of 147.6 GHz for synchrotron sources.  We arbitrarily selected PA2 as our standard for comparison of flux densities of compact sources from different arrays and used $3^\prime$ as a matching radius.  These checks of internal consistency, array by array,  are detailed in Appendix \ref{appendix:act_updates_note}.  With one exception, measurements of flux densities in each ACT band agree to 1-2\%.  The exception is PA3 operating at 144.1 GHz in season 2016, which recorded flux densities on average 7\% higher than PA2 for the same season and band. Hence we dropped PA3 season 2016 data at 144.1 GHz from further consideration.

\subsubsection{Consistency of ACT data from season to season}\label{subsubsec:act_s2s_consistency}

\begin{figure}
	\plotone{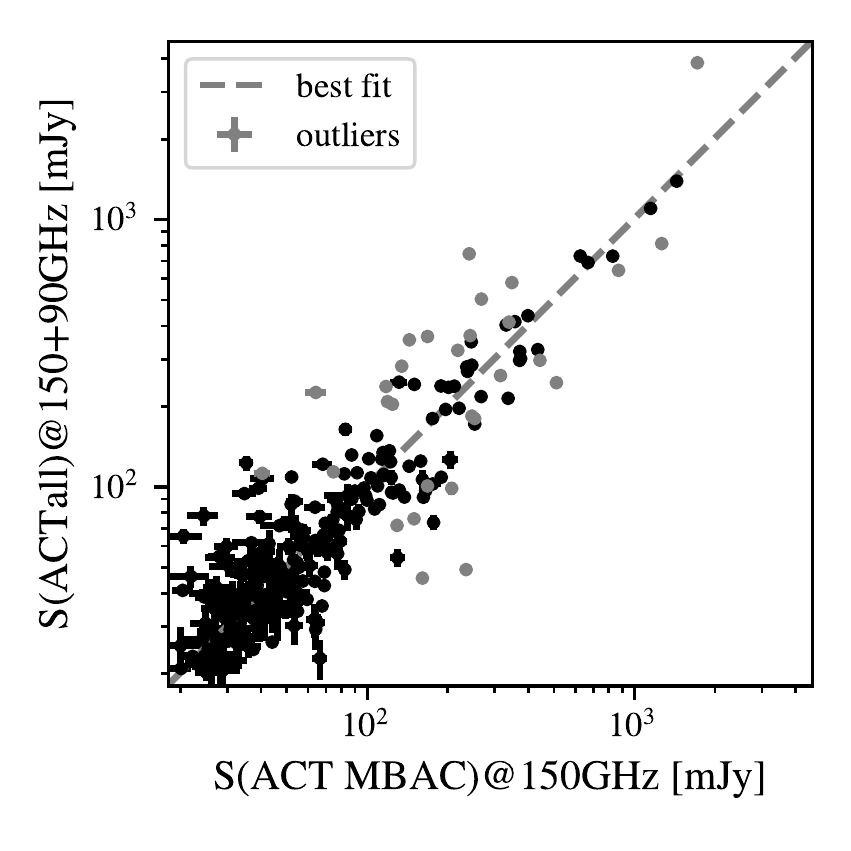}
	\caption{Comparing ACT measurements made with the MBAC array in 2008-10 with later (2013-2016) ACTPol measurements made at $\sim$145 GHz with polarized arrays (``ACTall"). We omit PA3 results from 2016 in the latter.  \added{The number of matches is 284}. The slope of the best-fit line (shown) is very close to unity at 0.9934.  Recall that the 30 outlier sources (grey) are not included in the fit and that we fit to linear data even though we show log-log plots. \label{fig:ACT_MBAC_v_ACTPol_avg}}
\end{figure}

Next, we turn to consistency of ACT measurements made in different seasons.  It is vital to the use of ACT as a bridge to connect Planck and ALMA measurements that we confirm the consistency of ACT flux densities across time, and that we measure any residual calibration changes accurately. Of particular importance is the comparison between ACT observations made in 2008-2010, when Planck was active, to the later observations (2013-16) when ALMA was operating.  It is this comparison that determines the quality of the bridge we build between Planck and ALMA flux densities.  Given the large span of time involved, we expect variability of the sources to play a prominent role.  To mitigate to some degree variability during each of the two periods 2008-10 and 2013-16, we compare weighted averages of flux densities made for each of the two periods.  For the later ACTPol observations (2013-16) specifically, we use inclusive catalogs of all 93 and $\sim$145 GHz sources (``ACTall”); the flux density for each is the inverse-variance weighted average of measurements across all arrays and all seasons 2013-16 (except PA3 in 2016 for reasons cited above).  The same is true for the earlier MBAC measurements made in 2008-10 \citep{marsden_atacama_2014,gralla_atacama_2020}.  Changes in luminosity of sources between the two epochs, of course, remain and are responsible for most of the scatter evident in Fig. \ref{fig:ACT_MBAC_v_ACTPol_avg}. In principle, given a large enough sample, source variability should not affect the slope of the fitted line, but only the scatter.  In Appendix \ref{appendix:var_bias} we reconsider the validity of this assumption in the case when the two catalogs to be compared have very different flux density thresholds, but that is not the case here.  To mitigate any remaining bias due to variability, and to decrease the scatter it introduces, we took all the steps described in \ref{subsec:flux_comp_general_features} above.

While the scatter is large, \added{due largely to source variability, }the average slope is extremely close to unity at $0.9934 \pm 0.0115 \pm (0.005)$ where the small \textit{additional} uncertainty introduced by sample variance is enclosed in parentheses.  We will continue to use this notation for other results. The newer ACTPol measurements are slightly smaller (by 0.7\%) than those made by MBAC in 2008-10.  The center frequency of the earlier MBAC observations was 147.6 GHz \citep{gralla_atacama_2020}, whereas the (color-corrected and averaged) center frequency of the later observations was close to 144.5 GHz.  Since most of our sources are synchrotron emitters with falling spectra, we could estimate the effect of these slightly different center frequencies: it is roughly 1-2\%. To make the small interpolation from 147.6 to 144.5 GHz, we employ spectral indices derived from ACT 93 and $\sim$145 GHz results, source by source, \textit{before} the fit shown in Fig. \ref{fig:ACT_MBAC_v_ACTPol_avg}.

We will also need to compare ACT observations made solely in 2016 (PA2 only; center frequency 144.7 GHz) with ACTall.  In this case, \added{there are over 5000 matches and} a much smaller correction for the difference in central frequencies is needed; for typical synchrotron spectra it is 0.1\%.  With this small correction included, we find $S(\text{PA2 season 2016})/S(\text{ACTall})= 1.0102 \pm 0.0030$. The full range of season to season comparisons and further details are provided in Appendix \ref{appendix:act_updates_note}.  We take these small differences into account when we use ACT observations as a bridge — see section \ref{subsec:linking_planck_alma}.

\subsubsection{Consistency of 93 and $\sim$145 GHz ACT data}\label{subsubsec:act_f2f_consistency}

Are ACT measurements at 93 and $\sim$145 GHz consistent?  Since we have only two bands, we cannot employ the same test of consistency used for Planck.  As a rough sanity check, we plotted 93 GHz flux densities measured in 2016 for $\sim$2000 sources against their 144.7 GHz flux densities.  The resulting slope was $1.302 \pm 0.002$, corresponding to an average spectral index of -0.60, a reasonable result.  For a subset of \added{400} bright sources, we also used Planck-derived spectral indices (based on 100 and 143 GHz measurements) to extrapolate the ACT 93 GHz values to 144.7 GHz.  There is obviously a great deal of variability, given that the Planck spectral indices were derived from observations made some 5 years earlier than the ACT ones.  Nevertheless, we find
$S(\text{93, extrapolated to 144.7 GHz})/S(144.7)
 = 0.962 \pm 0.006$.

These tests give us some confidence to use ACT 93 GHz observations to calculate ACT-based spectral indices.

\subsection{ Internal Consistency of SPT data}\label{subsec:spt_internal_consistence}

We also made use of the extensive catalog of compact sources released by the SPT collaboration \citep{everett_millimeter-wave_2020}.  Flux densities at $\sim$95, 150 and 220 GHz are provided; for observations of sources with synchrotron spectra, the effective center frequencies are closer to 97.43, 152.9 and 215.8 GHz (T. Crawford, private communication). Since three frequency bands are available, we may use the same procedure as for Planck to check frequency-to-frequency consistency.  For  \added{719} sources, we find that the flux densities measured at 152.9 GHz on average lie $1.39 \pm 0.25\%$ above the values predicted from the 97.43 and 215.8 GHz measurements.  This small discrepancy could be present because the SPT flux densities at 215.8 GHz are biased low, a point we return to below.

\section{Planck Based Calibration of ALMA, Using ACT Observations as a ``Bridge"}\label{sec:planck_based_calbration_ALMA}

Since we have measured the small, time-dependent changes in the calibration of both ACT and Planck, we can  include these small factors when comparing measurements of compact sources made by the two instruments (or by ALMA) no matter when they were made.  We expect, and find, that source variability dominates the scatter seen in any such comparison, despite the steps listed in \ref{subsec:flux_comp_general_features} to mitigate it.   

\subsection{Linking ALMA to Planck}\label{subsec:linking_planck_alma}

We outline here how we connect ALMA observations made in 2012 to 2017 with Planck observations made at an earlier epoch, 2009 August to 2012 January. We concentrate on frequencies between 90 and 110 GHz (and ALMA Band 3) at first.

\subsubsection{The ACT-Planck Comparison}\label{subsubsec:act_planck_comp}

The first step is to compare Planck observations made in 2009-10 with ACT observations made in a roughly similar time-frame, 2008-2010.  We thus compare Planck PCCS1 measurements (corrected to the 2018  calibration, as per section \ref{subsec:planck_internal_consistency}) to the ACT MBAC observations at 147.6 GHz (no 93 GHz ACT data were available until 2015) as  described by \citet{gralla_atacama_2020} and \citet{marsden_atacama_2014}; as noted, we take final flux density values from LAMBDA. The Planck 143 GHz data are color corrected and very slightly extrapolated to the MBAC center frequency using spectral indices derived from Planck data (section \ref{subsec:color_correction}). Specifically, we employ the weighted average of the spectral indices between 100 and 143 GHz and between 143 and 217 GHz to make this small extrapolation for each source. Since the 2008-2010 ACT observations covered only a small region of the sky, the number of matches is limited (only 33) and thus the uncertainty in this comparison is the dominant contributor to the uncertainty in the overall ALMA-Planck comparison.

\begin{figure}\label{fig:ACT_MBAC_v_Planck_PCCS1}
	\plotone{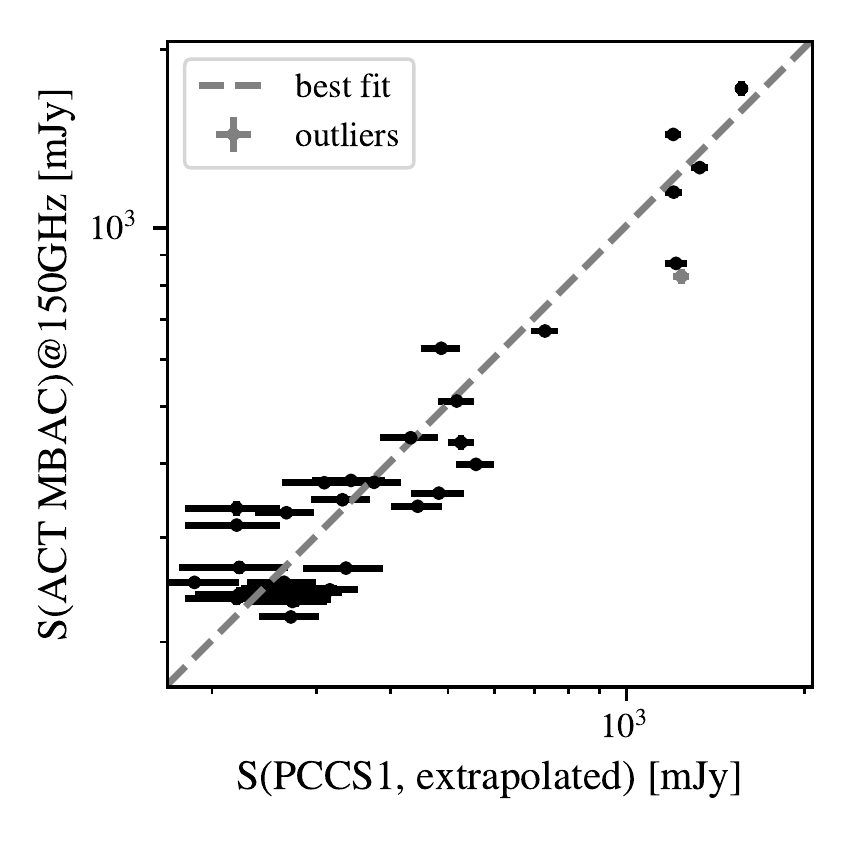}
	\caption{Comparing ACT flux densities \added{of 33 sources} measured by MBAC in 2008-2010 to those measured by Planck HFI at 143 GHz in 2009-10 (after color-correction and extrapolation to 147.6 GHz to match the center frequency of ACT’s MBAC array).  In this case, only a single iteration was required, with one source (shown in grey) excluded from the fit.  Note that the fit (dashed line) is extremely close to unity.}
\end{figure}

For this comparison, like all the others described in this paper, we take all the precautions described in \ref{subsec:flux_comp_general_features} to limit the effect of source variability.  One further step we take from the beginning is to limit the comparison to sources with flux densities S $\geq 220$ mJy in the ACT MBAC catalog.  This threshold was chosen as the apparent flux density level that matches the 90\% completeness in the 143 GHz Planck catalog.\footnote{The official 90\% completeness given in \citet{planck_collaboration_planck_2014-1} is 190 mJy, but we instead infer from figure 6 in that same paper that 220 mJy is a more conservative value, and we adopt it. Cutting at 190 mJy instead changes the slope of the fit by ${\sim}0.5 \sigma$.}  Adopting such a threshold ensures that there are not large numbers of faint ACT sources without matches in the Planck PCCS1 catalog (see Table \ref{table:PCCS1_v_MBAC_comp}).  

The two lists of sources being compared here have very different threshold sensitivities.  That introduces what we call ``variability bias,” a topic we address in more detail in Appendix \ref{appendix:var_bias}.  Briefly, a variable source that happened to be bright when Planck observed it, but faded during the ACT observations, would still be detected by ACT given its much greater sensitivity.  But the opposite is not true.  Indeed, we find three sources with ACT flux density above our chosen threshold of 220 mJy with no match in the Planck data, presumably because they fell below Planck’s detection limit in 2009-2011.  Note that this number (3 of 33) is consistent with what is expected given our use of the 90\% completeness limit, so their absence from the Planck catalog is not necessarily due to variability.  If we simply ignore these ``missing” matches, we might bias our results. Consequently, we replace the Planck flux density of these three sources by either $320$ mJy, or $190$ mJy, values chosen as reasonable limits on the possible flux density of the three sources not found in the Planck catalog (see Appendix \ref{appendix:var_bias}). We assign an uncertainty of $\pm40$ mJy to these three replacements, a value typical of Planck flux density uncertainties.  The resulting slopes for the two different limits are in close agreement at $1.0251 \pm 0.0258 \pm (0.0150)$ and $1.0176 \pm 0.0308 \pm (0.0140)$.  We tested this procedure by (a) simply omitting the three sources not detected by Planck, changing the slope by ${\sim}0.5\sigma$; and (b) replacing the three sources not found in Planck with an intermediate value $S = 220 \pm 40$ mJy instead, changing the slope to $1.0149 \pm 0.0298 \pm (0.0161)$.  We also tried lowering the detection threshold from 220 to 190 mJy; we found more matches (39), but also twice as many ACT sources with no match in the Planck catalog. Replacing more unmatched sources changed the slope by ${\sim}0.5\sigma$.  These and other tests are described in more detail in Appendix \ref{appendix:var_bias}.  For the remainder of this work, we adopt a value lying between the two slopes set by plausible limits on the flux density of the 3 missing Planck sources.  Taking the inverse-variance weighted average, we find ACT MBAC fluxes on average = $1.0220 \pm 0.0283$ those from Planck. Here we have assumed complete correlation in the errors.

\added{Given the small number of sources available for this comparison, we also calculated the sample variances using the bootstrap method described in section \ref{subsubsec:sample_variance}.   As expected, they were larger than the statistical errors.  For the two fits just described, the bootstrap estimates of the errors are 0.0408 and 0.0448.  We take the inverse variance weighted average of the mean and uncertainties to report S(MBAC)/S(Planck) = $1.0220 \pm 0.0283 \pm (0.0148)$. In this case, the statistical error estimate for our fit was $\pm 0.0283$, and the bootstrap estimate was $\pm 0.0431$, leading us to estimate the additional contribution from sample variance to be $\pm0.0148$. }

Thus, after making the small corrections necessary to bring PCCS1 calibration into line with PCCS2 values (Table \ref{table:calibration_table}), we find good agreement between Planck and MBAC flux density measurements for sources with 143 GHz flux density above 220 mJy.  For the moment, we retain ``excess" significant figure in these results to allow more precise combinations with other results.  

One additional step is needed, however: from Table \ref{table:PCCS1_v_PCCS2_comp} we see that even after applying the published flux density corrections, on average corrected PCCS1 flux densities run (0.48 $\pm$ 0.33)\% below PCCS2 values, so we include that small correction and propagate the errors appropriately. Finally,we obtain 

\begin{equation}\label{eq:MBAC_v_Plack_slope_final}
\text{S(MBAC)/S(Planck)} = 1.0171 \pm 0.0284 \pm (0.0148).
\end{equation}

\begin{deluxetable*}{llDDDcDDcDDcD}
	\tablecaption{Flux densities from Planck PCCS1 at 143 GHz, both before and after color correction and extrapolation to 147.6 GHz.  We take PCCS2 and the 2018 Planck calibration as the standard, and hence have corrected the PCCS1 data by the small factors given in Table \ref{table:calibration_table} before entering them in col. 5. Values in col. 6 have been color corrected and extrapolated as well.  These may be compared to ACT MBAC measurements for the same sources (col.7). \added{One MBAC source was inadvertently omitted from LAMBDA and hence has no official ACT name in col. 2.}
	\label{table:PCCS1_v_MBAC_comp}}
	
	\tablehead{
		\colhead{Planck ID} &\colhead{ACT ID}&\multicolumn2c{RA} &\multicolumn2c{DEC} &\multicolumn5c{Raw Planck Flux\tablenotemark{a},}&\multicolumn5c{Corrected/Extrapolated}&\multicolumn5c{MBAC Flux, mJy}\\
		\colhead{} &\colhead{}&\multicolumn2c{} &\multicolumn2c{} &\multicolumn5c{mJy}&\multicolumn5c{Planck Flux, mJy}&
	}
	\decimals
	\startdata
		PCCS1 G115.22-64.77 & ACT-S J003820-020738 & 9.58   & -2.10  & 438  & $\pm$ & 48 & 433  & $\pm$ & 48 & 442.5  & $\pm$ & 3.2  \\
		PCCS1 G117.03-61.30 & ACT-S J004013+012542 & 10.03  & 1.44   & 185  & $\pm$ & 35 & 187  & $\pm$ & 35 & 252.2  & $\pm$ & 2.0  \\
		PCCS1 G134.13-60.01 & ACT-S J011343+022211 & 18.44  & 2.37   & 344  & $\pm$ & 48 & 343  & $\pm$ & 48 & 374.8  & $\pm$ & 6.2  \\
		PCCS1 G141.08-61.74 & ACT-S J012528-000556 & 21.34  & -0.07  & 451  & $\pm$ & 45 & 445  & $\pm$ & 44 & 339.0  & $\pm$ & 2.1  \\
		PCCS1 G288.24-63.93 & ACT-S J013306-520006 & 23.28  & -51.99 & 223  & $\pm$ & 35 & 222  & $\pm$ & 35 & 240.6  & $\pm$ & 7.5  \\
		PCCS1 G276.06-61.77 & ACT-S J021046-510103 & 32.72  & -51.01 & 1584 & $\pm$ & 39 & 1565 & $\pm$ & 38 & 1718.3 & $\pm$ & 50.0 \\
		PCCS1 G162.13-54.41 & ACT-S J021749+014446 & 34.45  & 1.75   & 1222 & $\pm$ & 53 & 1213 & $\pm$ & 52 & 871.2  & $\pm$ & 2.8  \\
		PCCS1 G272.51-54.59 & ACT-S J025329-544152 & 43.38  & -54.72 & 1261 & $\pm$ & 39 & 1239 & $\pm$ & 38 & 828.7  & $\pm$ & 24.1 \\
		PCCS1 G262.04-31.84 & ACT-S J054045-541821 & 85.22  & -54.30 & 538  & $\pm$ & 27 & 526  & $\pm$ & 27 & 434.4  & $\pm$ & 13.1 \\
		PCCS1 G216.98+11.38 & ACT-S J073918+013658 & 114.83 & 1.63   & 1201 & $\pm$ & 39 & 1200 & $\pm$ & 39 & 1438.0 & $\pm$ & 3.6  \\
		PCCS1 G219.89+11.73 & ACT-S J074554-004420 & 116.46 & -0.74  & 316  & $\pm$ & 48 & 309  & $\pm$ & 47 & 371.5  & $\pm$ & 3.8  \\
		PCCS1 G220.71+18.57 & ACT-S J081126+014648 & 122.87 & 1.78   & 488  & $\pm$ & 38 & 487  & $\pm$ & 38 & 626.4  & $\pm$ & 5.4  \\
		PCCS1 G221.08+19.49 & ACT-S J081523+015455 & 123.85 & 1.90   & 278  & $\pm$ & 41 & 277  & $\pm$ & 41 & 235.6  & $\pm$ & 6.2  \\
		PCCS1 G338.74+60.07 & ACT-S J135927+015954 & 209.84 & 2.01   & 227  & $\pm$ & 47 & 222  & $\pm$ & 46 & 267.3  & $\pm$ & 6.6  \\
		PCCS1 G343.15+58.63 & ACT-S J141004+020305 & 212.51 & 2.04   & 525  & $\pm$ & 36 & 517  & $\pm$ & 36 & 510.3  & $\pm$ & 6.8  \\
		PCCS1 G001.34+45.98 & ACT-S J151640+001500 & 229.16 & 0.23   & 733  & $\pm$ & 39 & 729  & $\pm$ & 39 & 669.9  & $\pm$ & 4.7  \\
		PCCS1 G006.80+43.23 & ACT-S J153452+013101 & 233.73 & 1.54   & 494  & $\pm$ & 51 & 483  & $\pm$ & 50 & 356.7  & $\pm$ & 4.1  \\
		PCCS1 G009.58+37.70 & ACT-S J155751-000151 & 239.45 & -0.01  & 284  & $\pm$ & 48 & 278  & $\pm$ & 47 & 246.1  & $\pm$ & 3.9  \\
		PCCS1 G019.54+27.24 & ACT-S J165103+012917 & 252.76 & 1.47   & 337  & $\pm$ & 52 & 336  & $\pm$ & 52 & 266.6  & $\pm$ & 3.0  \\
		PCCS1 G052.40-36.50 & ACT-S J213410-015320 & 323.55 & -1.88  & 1342 & $\pm$ & 43 & 1331 & $\pm$ & 43 & 1264.0 & $\pm$ & 3.1  \\
		PCCS1 G055.46-35.57 &                      & 324.15 & 0.70   & 1239 & $\pm$ & 39 & 1202 & $\pm$ & 38 & 1149.0 & $\pm$ & 3.0  \\
		PCCS1 G057.68-40.34 & ACT-S J215614-003705 & 329.06 & -0.64  & 381  & $\pm$ & 42 & 375  & $\pm$ & 42 & 372.1  & $\pm$ & 2.4  \\
		PCCS1 G059.87-42.37 & ACT-S J220643-003103 & 331.66 & -0.54  & 273  & $\pm$ & 32 & 272  & $\pm$ & 32 & 220.5  & $\pm$ & 2.4  \\
		PCCS1 G065.83-45.32 & ACT-S J222646+005209 & 336.68 & 0.90   & 322  & $\pm$ & 37 & 316  & $\pm$ & 36 & 245.0  & $\pm$ & 2.5  \\
		PCCS1 G070.17-49.71 & ACT-S J224730+000005 & 341.88 & 0.02   & 271  & $\pm$ & 31 & 267  & $\pm$ & 31 & 330.6  & $\pm$ & 2.4  \\
		PCCS1 G072.93-50.22 & ACT-S J225404+005419 & 343.51 & 0.89   & 266  & $\pm$ & 35 & 265  & $\pm$ & 35 & 252.3  & $\pm$ & 2.4  \\
		PCCS1 G071.86-53.50 & ACT-S J230108-015804 & 345.28 & -1.95  & 336  & $\pm$ & 38 & 332  & $\pm$ & 38 & 347.7  & $\pm$ & 2.9  \\
		PCCS1 G079.31-52.08 & ACT-S J231101+020502 & 347.76 & 2.08   & 283  & $\pm$ & 40 & 273  & $\pm$ & 40 & 234.2  & $\pm$ & 3.4  \\
		PCCS1 G080.46-57.74 & ACT-S J232653-020211 & 351.73 & -2.02  & 264  & $\pm$ & 41 & 265  & $\pm$ & 41 & 246.3  & $\pm$ & 2.9  \\
		PCCS1 G084.24-58.55 & ACT-S J233520-013111 & 353.82 & -1.53  & 296  & $\pm$ & 44 & 288  & $\pm$ & 43 & 242.6  & $\pm$ & 3.1  \\
		None                & ACT-S J005905+000652 & 14.77  & 0.11   & .    &       & .  & .    &       & .  & 315.1  & $\pm$ & 2.2  \\
		None                & ACT-S J022912-540325 & 37.30  & -54.06 & .    &       & .  & .    &       & .  & 336.6  & $\pm$ & 10.1 \\
		None                & ACT-S J151216+020318 & 228.07 & 2.06   & .    &       & .  & .    &       & .  & 237.1  & $\pm$ & 5.8 
	\enddata
	\tablenotetext{a}{Calibration corrected to 2018 level}
\end{deluxetable*}

Since ACT observations at the lower frequency of 93 GHz began only in 2015, we have no simultaneous Planck data with which to compare them.  Instead, we simply compare ACT 93 GHz flux densities from 2016 with those from the Planck 100 GHz PCCS2 catalog, after color-correcting and extrapolating the latter using Planck-based spectral indices.  Since the approximate 90\% completeness limit at 100 GHz is higher than for 143 GHz, we employ a threshold of 330 mJy in the comparison, and replace any ACT sources not found in Planck with a Planck flux density set to 400 $\pm$ 50 mJy (the uncertainty roughly matches typical Planck values at 100 GHz).  At 93 GHz, we find
\begin{equation}\label{eq:ACT93_v_Plack_slope_final}
	\text{S(ACT 93 GHz)/S(Planck)} = 1.0403 \pm 0.0269 \pm (0.0093).
\end{equation}
We have added 0.13\% Planck beam uncertainty to the error listed. Including a small number of 93 GHz observations made in 2015 changes the slope by ${\sim}0.1\sigma$.  \added{Not replacing the "missing" sources increases the slope by ${\sim}0.25\sigma$.} We return to the evidence that ACT flux densities at 93 GHz may be slightly overestimated in section \ref{subsubsec:using_93GHz_act}.  \new[]{Recall that any small \added{constant or multiplicative} bias in ACT values will cancel out to a large degree, since we compare ACT to both Planck and ALMA.}

\subsubsection{Resolution differences}\label{subsubsec:res_differences}

The solid angle of Planck beams is at least an order of magnitude larger than those of ACT or SPT; we need to check whether Planck measurements of compact sources could be biased high by background sources in the Planck beam. While the DETFLUX method used to determine Planck flux densities listed in PCCS2 averages out the contributions from a uniform background of other sources, sources preferentially clustered around a Planck source could bias Planck measurements high. \citet{welikala_probing_2016} present evidence that Planck's large beam is picking up flux from star-forming galaxies along the line of sight to strongly lensed dusty star forming galaxies.  On the other hand, that same paper shows no such effect for (unlensed) synchrotron sources like those examined here.  In addition, the amplitude of the signal observed for the lensed sources is at most a few mJy, $<1\%$ of the flux density of Planck sources we consider.  

We also estimated the probability of finding a background source brighter than either 22 or 4.4 mJy within the $6^\prime$ search radius, assuming those sources are randomly placed on the sky.  We selected these two limits since they represent $\sim$10\% and $\sim$2\% of the flux density of the faintest sources we use in the Planck-ACT comparison.  To make these estimates, we used the 150 GHz source counts of \citet{everett_millimeter-wave_2020}; ACT data, published by \citet{datta_atacama_2019} are in general agreement with the SPT source counts, but the format of the SPT counts makes them easier to use.  The probability of finding a source with $S > 22$ mJy inside a randomly placed circle of radius $6^\prime$ is less than 4\%, so we might expect to find one or two such random background sources among the 33 Planck-ACT matches.  The probability for finding a weaker (4.4 mJy) source is larger, 24\%, so we expect several such cases.

We can use the ACT MBAC catalog, with no restriction on flux density, to record the number of secondary ACT sources we actually observed within $6^\prime$ of each Planck source.  Ten of the 33 sources plotted on Fig. \ref{fig:ACT_MBAC_v_Planck_PCCS1} do have one or more secondary sources nearby; in only 3 cases, does the flux density of a secondary source exceed 22 mJy.  Neither value is strongly inconsistent with the expectations for randomly placed sources

Planck's broad beam (with a full width at half maximum of ${\sim}7.2^\prime$ at 150 GHz) will incorporate some of the emission from these secondary sources.  How much each secondary source contributes depends on its distance from the primary source.  If, for instance, the secondary source lies $3.6^\prime$ away, only $\frac{1}{2}$ of its flux density is included in the Planck measurements.  When this convolution with the Planck beam is taken into account, we find that the observed secondary sources can contribute $\sim$1\% to  $\sim$10\% of the total flux density measured by Planck for the 11 cases where there are secondary sources; the other 22 sources are unaffected. 

If we reduced the Planck flux densities by these small amounts, using a rough model for the Planck beam, and then repeat the fit shown in Fig. 3, we find a change in the slope of only $\sim$0.5\%, or $\sigma/6$.  The matched filter used in Planck source extraction would reduce this still further; hence we continue to use the value shown in Eq. \ref{eq:MBAC_v_Plack_slope_final}.

\subsubsection{Using ACT 147.5 GHz observations as a bridge to ALMA}\label{subsubsec:act_bridge2ALMA}

The next step is to use additional ACT measurements of extragalactic sources as a bridge.  We begin with the comparison of 2008-10 MBAC measurements to the all-season, all-array catalog of DR4 ACTPol measurements made in 2013-16 (``ACTall"; recall that this omits $\sim$145 GHz 2016 data from PA3). Here, of course, the variability of sources introduces evident scatter in the fit of Fig. \ref{fig:ACT_MBAC_v_ACTPol_avg}.  On the other hand, we apply a lower threshold of 20 mJy, chosen to be ${\sim}10\sigma$, to eliminate many weak sources.  The lower flux density cut allows many more (284) matches between the two ACT data sets.  The result is that the error is smaller than for the earlier MBAC-Planck comparison.  From section \ref{subsubsec:act_s2s_consistency}, we take $S(\text{ACTall})/S(\text{MBAC}) =0.9934 \pm 0.0114 \pm (0.005)$.  The second leg of the bridge is to compare ACT measurements made in 2016 (PA2 only) with ACTall.  In this case, a much smaller adjustment for frequency differences is needed.  From the same section, we find $S(\text{PA2 season 2016})/S(\text{ACTall})= 1.0102 \pm 0.0030$.

Note that since both MBAC and 2016 PA2 data are compared to ACTall data, small uncertainties in the latter cancel out to some degree.  Combining these results with Eq. \ref{eq:MBAC_v_Plack_slope_final}, we thus have tied the 2016 PA2 observations made at 144.7 GHz to Planck, using MBAC and ACTall as bridges:
\begin{equation}\label{eq:S16PA2_v_Planck}
	\text{S(ACT S16)/S(Planck)} = 1.0207\pm0.0309\pm(0.0158),
\end{equation}
where we have summed the uncertainties from all previous steps in quadrature.  Note that, for now, we still retain ``excess" significant figures to enable precise combinations of the various results.  ACT measurements of compact sources at ${\sim}145$ GHz match Planck's within errors, but may run a bit high.  \added{Recall that a small multiplicative offset will cancel out, since we also compare ACT to ALMA.}  To test this finding, we compare extrapolated and color-corrected PCCS2 measurements at 143 GHz directly with ACTall.  Since the two sets of observations do not overlap in time, variability increases the scatter, but we have many more sources (291) in common. We find $\text{S(ACTall)/S(Planck)} = 1.0332 \pm 0.0241 \pm (0.0117)$, in excellent agreement with the result above.

\begin{figure}
    \plotone{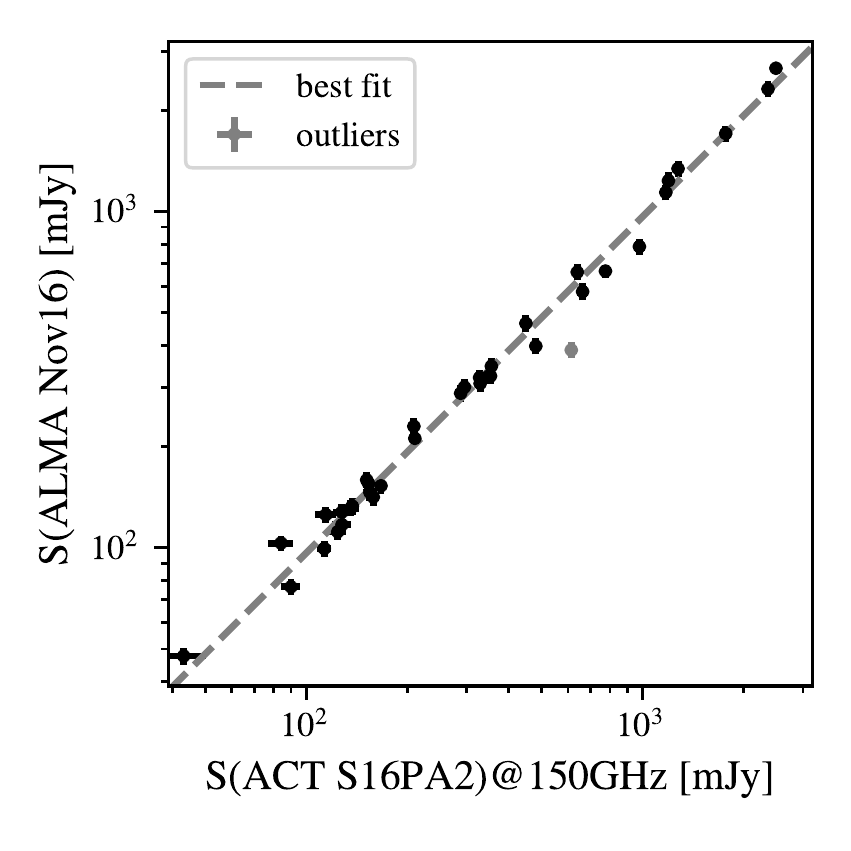}
	\caption{Dedicated Nov. 2016 ALMA measurements \added{of 37 sources} compared to ACT measurements in the same year to reduce the effect of source variability (note the reduced scatter compared to Figs. 2 and 3).  ALMA-based spectral indices were used to interpolate the ALMA fluxes to 144.7 GHz to match ACT.  Only one outlier source (shown in grey) was dropped from this fit; this source, J2236-1433, is known to vary over short time scales. \label{fig:ACT_S16PA2_v_ALMA_Nov2016}}
\end{figure}

Finally, we compare the PA2 2016 ACT values to the dedicated ALMA measurements made in November during the ACT 2016 season (see section \ref{subsec:ALMA_flux_densities}).  Here, we use ALMA measurements at 343.5, 91.5 and 103.5 GHz to interpolate 37 ALMA measurements to $\sim$145 GHz.

\begin{equation}\label{eq:PA2S16_v_ALMA_Nov16}
	\text{S(ALMA)/S(ACT S16)} = 0.9654 \pm 0.0144 \pm (0.0037),     
\end{equation}
as shown in Fig. \ref{fig:ACT_S16PA2_v_ALMA_Nov2016}, and consequently,  
\begin{equation}\label{eq:ALMA_v_Planck_slope_raw}
	\text{S(ALMA)/S(Planck)} = 0.9854\pm0.0333\pm(0.0155),        
\end{equation}
thus linking ALMA calibration in band 3 to the absolute calibration of the Planck satellite.  Calibration of the two instruments agrees well within errors (which are dominated by the first step in the process just described, the MBAC vs. Planck comparison).

We can eliminate one step in the bridge, namely the use of the multi-season ACTall data, by comparing the MBAC measurements to both Planck and the later season 2016 PA2 ACT measurements, then comparing the latter to ALMA.  Omitting ACTall, we find $S(\text{ALMA})/S(\text{Planck}) = 0.9055\pm0.0331\pm(0.0164)$. The problem here is the long gap between 2008-10 MBAC measurements and season 2016.  We may also drop the early MBAC observations and link Planck to ACTall, then ACTall to the season 2016 PA2 measurements and hence to ALMA: $S(\text{ALMA})/S(\text{Planck}) = 1.0076\pm0.0280\pm(0.0117)$.  Eliminating season 2016 PA2 instead, by linking ACTall directly to ALMA and through MBAC to Planck, yields marginally consistent results but larger error: $S(\text{ALMA})/S(\text{Planck}) = 0.9519\pm0.0393\pm(0.0139)$. We may also drop both MBAC and season 2016 PA2, and use ACTall as the sole link between Planck and ALMA, yielding $S(\text{ALMA})/S(\text{Planck})= 0.9734\pm0.0355\pm(0.0104)$.  Finally, we can simply compare the ACT PA2 measurements made in 2016 with the corrected PCCS2 Planck data to move directly to the equivalent of Eq. \ref{eq:S16PA2_v_Planck}, after adding the $0.07\% $ uncertainty in the Planck beam solid angle to the error:
\begin{equation}
	S(\text{PA2 S16})/S(\text{Planck}) = 1.0245 \pm 0.0241 \pm (0.0115),
\end{equation}
in excellent agreement with the result found earlier.  Combining this result with Eq. \ref{eq:PA2S16_v_ALMA_Nov16} yields
\begin{equation}\label{eq:ALMA_v_Planck_omit_MBAC_ACTall}
	S(\text{ALMA})/S(\text{Planck}) = 0.9890\pm0.0276\pm(0.0115),
\end{equation}
consistent with Eq. \ref{eq:ALMA_v_Planck_slope_raw}.  All but one of the ALMA-Planck comparisons are consistent with unity within the error bars. With that same exception, they are internally consistent within the errors as well.

\subsubsection{Using ACT 93 GHz observations}\label{subsubsec:using_93GHz_act}

We can repeat the last step using 2016 ACT data at 93 GHz, comparing it to the extrapolated and color-corrected Planck 100 GHz data, as well as to ALMA data.  From section \ref{subsec:linking_planck_alma}, we have $S(\text{ACT 93 GHz})/S(\text{Planck}) = 1.0403 \pm 0.0269 \pm (0.0093)$. Next, as we did earlier for the higher ACT frequency, we compare 32 ACT 93 GHz measurements made in 2016 to ALMA 91.5 and 103.5 GHz measurements, using ALMA spectral indices to interpolate to 93 GHz. This gives
\begin{equation}
	\text{S(ALMA)/S(ACT 93GHz)} = 0.9687 \pm 0.0133 \pm (0.0028),
\end{equation}
again suggesting that ACT 93 GHz values may be a few percent high.  Combining this result with Eq. \ref{eq:ACT93_v_Plack_slope_final}, we find:

\begin{equation}\label{eq:ACT_v_Plack_slope_90GHz}
	\text{S(ALMA)/S(Planck)} = 1.0077\pm0.0295\pm(0.0093) .
\end{equation}
Since both Planck and ALMA are compared to the 93 GHz ACT data, small errors in the latter cancel to some degree.  Employing entirely different Planck and ACT catalogs produces results in acceptable agreement with those found in Eq. \ref{eq:ALMA_v_Planck_slope_raw}, though somewhat larger.

\subsubsection{Using ACT not ALMA spectral indices} \label{subsubsec:act_alma_comp_w_act_alpha}

A potential problem with the process just outlined is our reliance on spectral indices from ALMA observations at widely separated frequencies (${\sim}100$ and ${\sim}340$ GHz).  We can instead use spectral indices for each source derived from the 2016 ACT measurements at 93 and $\sim$145 GHz to interpolate the ALMA 91.5 and 103.5 GHz measurements to ACT’s central frequency of 144.7 GHz.  This of course assumes that the 93 GHz ACT data is as well calibrated as the higher frequency data.  If the ACT 93 GHz flux densities are on average a bit high (as suggested by the comparison with Planck and ALMA), employing them here will bias the ACT measurements high. Using ACT-based spectral indices we find, in place of Eq. \ref{eq:PA2S16_v_ALMA_Nov16}, 
\begin{eqnarray}
S(\text{ALMA 91.5 GHz})&/&S(\text{ACT S16})\label{eq:ALMA91_v_ACT_slope}\\ &=& 0.9610 \pm 0.0122 \pm (0.0019),\nonumber\\
S(\text{ALMA 103.5 GHz})&/&S(\text{ACT S16})\label{eq:ALMA103_v_ACT_slope}\\ &=& 0.9661 \pm 0.0108 \pm (0.0012).\nonumber
\end{eqnarray}

\replaced{We simply take the inverse variance weighted average of these to obtain 
\begin{equation}\label{eq:ALMA_v_ACT_avg_w_ACT_alpha}
	S(\text{ALMA})/S(\text{ACT S16}) = 0.9638 \pm 0.0115 \pm (0.0015),
\end{equation}

\new[Then, as before, we use ACT results as a bridge to link ALMA to Planck, employing Eq. \ref{eq:S16PA2_v_Planck}:]{To reach this result, we have used both 93 and $\sim$145 GHz data from ACT 2016 observations.  At both frequencies, ACT measurements are slightly higher than Planck's (Eqn. \ref{eq:ACT93_v_Plack_slope_final} and \ref{eq:S16PA2_v_Planck}), but with different ratios.  We take the inverse variance weighted average of two: $S(\text{ACT S16})/S(\text{Planck}) = 1.0317 \pm 0.0292 \pm (0.XXX)$, which differs from either value by $\sim$0.4 sigma.  Combining that with Eq. \ref{eq:ALMA_v_ACT_avg_w_ACT_alpha}, we find}
\begin{equation}
	S(\text{ALMA})/S(\text{Planck}) = 0.9948 \pm 0.0313
\end{equation}
in good agreement with most preceding results.}{To make the connection to Planck we need to compare ACT to Planck at the appropriate ALMA  frequencies. To do this we interpolate the PCCS2 flux densities to 91.5 GHz and 103.5GHz using Planck based spectral indices and compare them to ACT measurements that have been extrapolated to these frequencies using ACT spectral indices. At 91.5 GHz we find $S(\text{ACT S16})/S(\text{Planck}) = 1.0486\pm0.0321\pm(0.0155)$ (with $N=147(7)$) and at 103.5GHz $S(\text{ACT S16})/S(\text{Planck})=1.0436\pm0.0329\pm(0.0134)$ ($N=167(7)$). In both cases we added the 0.13\% beam solid angle error appropriate for the Planck 100GHz data; we also required $S>330$ as in the case for the 93 GHz comparison discussed above.}  \added{Using these results we find
\begin{eqnarray}
S(\text{ALMA 91.5 GHz})&/&S(\text{Planck})\label{eq:ALMA91_v_Planck_slope}\\ &=& 1.0077\pm0.0334\pm(0.0147),\nonumber\\
S(\text{ALMA 103.5 GHz})&/&S(\text{Planck})\label{eq:ALMA103_v_Planck_slope}\\ &=& 1.0082\pm0.0338\pm(0.0127)\nonumber
\end{eqnarray}
in good agreement with preceding results.}

The results of similar calculations for other ACT and ALMA data sets are provided in the summary table in section \ref{sec:discussion} below.  Overall, we find that the Uranus-based calibration of ALMA in band 3 is in reasonable agreement with the absolute calibration of Planck.

\subsection{Including other ALMA observations}\label{subsec:including_other_alma_obs}

We now turn to ALMA observations from the larger list of ALMA observations from cone searches and grid monitoring, as described in section \ref{subsec:ALMA_flux_densities}.  As noted there, these are of mixed quality and cover a wider range of time.  Source variability was therefore a more significant problem.  To minimize it, we began by separating the list of observations by year, retaining only observations made in late 2015 and 2016.  These were compared to ACT measurements made in the same 2016 season.  To calculate ALMA spectral indices, we required sources with nearly simultaneous observations in a least two different ALMA bands, one of them Band 3.  As in \ref{subsec:linking_planck_alma}, these were used to interpolate (or extrapolate) ALMA observations at various frequencies to ACT’s central frequency of $\sim145$ GHz.

After these cuts, we found 96 sources with valid ALMA observations, including the dedicated November observations; we then compared these measurements to those made by ACT in the 2016 season, using ALMA-based spectral indices.  \added{This provides a useful check on the work reported above.}  As we did above (Eq. \ref{eq:S16PA2_v_Planck}), using Planck-based spectral indices, we also compare ACT to Planck.  For this wider set of ALMA observations, we find:
\begin{equation}\label{eq:ALMAall_v_Planck}
	S(\text{ALMA})/S(\text{Planck})= 1.0006 \pm 0.0349  \pm (0.0146),
\end{equation}
fully consistent with the earlier result (Eq. \ref{eq:ALMA_v_Planck_slope_raw}).  We may also omit various portions of the ``bridge" linking Planck to ALMA, as we did in section \ref{subsubsec:act_bridge2ALMA}.  For instance, if we use ACTall as the sole link from Planck to ALMA, we now obtain $0.968 \pm 0.034 \pm (0.012)$; using season 2016 PA2 data as the link instead yields $1.004 \pm 0.029 \pm (0.010)$.  

We also found a handful of additional sources among the list of bright quasars used for the grid observations (see section \ref{subsec:ALMA_flux_densities}) at various times in late 2015 or 2016.  Including these in the fit changed the slope of the fits by less than $0.5 \sigma$, except for the case where we use ACTall as the sole bridge between Planck and ALMA (here, adding the grid observations raised the slope by ${\sim} 1 \sigma$ to a value closer to unity).

Using ACT 93 GHz observations, as in section \ref{subsubsec:using_93GHz_act}, as a ``bridge" between Planck 100 GHz observations and the wider set of ALMA observations yields $1.026 \pm 0.030 \pm (0.010)$.

Had we used ACT-based spectral indices instead (as in section \ref{subsubsec:act_alma_comp_w_act_alpha}), we would have
\begin{eqnarray}
S(\text{ALMA 91.5 GHz})&/&S(\text{Planck})\label{eq:ALMAall91_v_Planck_slope}\\ &=& 1.0266\pm0.0343\pm(0.0156)\nonumber\\
S(\text{ALMA 103.5 GHz})&/&S(\text{Planck})\label{eq:ALMAall103_v_Planck_slope}\\ &=& 1.0207\pm0.0354\pm(0.0126),\nonumber
\end{eqnarray}
instead; \added{slightly higher, but consistent well within the uncertainty}. In Table \ref{table:summary_results} we summarize the many comparisons we make between Planck and ALMA results.  Note that all the ALMA/Planck comparisons using the expanded ALMA data set are consistent both with unity and with the results obtained using the dedicated ALMA observations only.



\subsection{SPT}\label{subsec:SPT_comp}

  SPT measurements of extragalactic sources were made in the interval 2008-11 \citep{everett_millimeter-wave_2020}, and hence roughly overlap with the Planck mission. We can thus compare them directly  to Planck observations.  As we did for ACT data, we apply all the measures from section \ref{subsec:flux_comp_general_features} to limit the effects of variability, color correct and extrapolate the Planck data, and set thresholds of  $S>330$ mJy at 90 GHz and $S>220$ mJy for 150 GHz.  The agreement between 97.43 GHz flux densities from SPT and Planck 100 GHz measurements is good, as is the agreement between 152.9 and Planck 143:
\begin{eqnarray}
	S(\text{SPT 97.43 GHz})&/&S(\text{Planck})\label{eq:SPT97_v_PCCS2}\\ &=& 1.0139 \pm 0.0269 \pm (0.0029)\nonumber \\
	S(\text{SPT 152.9 GHz})&/&S(\text{Planck})\label{eq:SPT152_v_PCCS2}\\ &=& 0.9639 \pm 0.0265 \pm (0.0036)\nonumber
\end{eqnarray}

At the highest frequency, the match is less good; 
\begin{equation}\label{eq:SPT215_v_PCCS2}
\begin{split}
S(\text{SPT 215.8 GHz})/&S(\text{Planck})\\ =& 0.8441 \pm 0.0408 \pm (0.0026).
\end{split}
\end{equation}

We employed Planck-based spectral indices to make the small adjustments for the differences between SPT and Planck frequencies.  Since there are three SPT frequencies, we could have used SPT-based spectral indices instead: this would have changed the results by ${\sim}0.3\sigma$ or less. 
Note that SPT values at ${\sim}150$ GHz run ${\sim}1\sigma$ below Planck; for ACT at $\sim$145 GHz, we earlier found values in close agreement or ${\sim}1\sigma$ above Planck. The SPT measurements at 215.8 GHz fall significantly below Planck 217 GHz values; a miscalibration could explain the small frequency-to-frequency discrepancy noted in section \ref{subsec:spt_internal_consistence}.

The optics of SPT are a better match to ACT than Planck.  In addition, a portion of the ACT fields overlaps the SPT survey area.  So we compared SPT measurements at 90 and 150 GHz with season 2016 ACT values, in this case using a threshold of 20 mJy for both catalogs.  After correcting for the different central frequencies of the two experiments, using ACT-based spectral indices, we find:
\begin{eqnarray}
	S(\text{ACT S16})&/&S(\text{SPT 97.43GHz})\\&=& 0.9849 \pm 0.0163 \pm (0.0065),\nonumber\\
    S(\text{ACT S16})&/&S(\text{SPT 152.9GHz})\\&=& 0.9286 \pm 0.0184  \pm (0.0082).\nonumber
\end{eqnarray}
Using SPT-based spectral indices to correct for the difference in central frequencies produces results consistent with the above within ${\sim}1\sigma$.  It may appear surprising that we find some SPT flux densities lower than Planck, but above ACT while ACT matches Planck fairly well.  This is possible because very different sets of sources are involved in these various comparisons (e.g., only bright sources when Planck is involved).  Given the reasonable match between SPT and Planck at 97.43 GHz, we tried using SPT observations as \replaced{a link}{ the first step} in the bridge between Planck and ALMA; \added{we connect Planck to SPT, SPT to ACT S16 at the appropriate SPT center frequencies and finally compare ALMA at those frequencies}. We find
$S(\text{ALMA})/S(\text{Planck}) = 0.9676\pm0.0324\pm(0.0061)$, consistent with previous results though somewhat lower.  At 152.9 GHz instead, we find $S(\text{ALMA})/S(\text{Planck}) = 0.8452\pm0.0322\pm(0.0071)$, and do not consider this result further.

\section{Discussion}\label{sec:discussion}

\begin{deluxetable*}{ccDcDcDl}[t]
	\tablecaption{Comparing ALMA flux density measurements to those from Planck, for various different combinations of the data.  Most of the entries come from the text (as indicated in column 1).  Entries in parentheses in column 2 are based on the more extensive set of ALMA measurements described in section \ref{subsec:including_other_alma_obs}.\label{table:summary_results}}
	\tablewidth{0pt}
	\tablehead{
	 &	\colhead{See in text} & \multicolumn8c{$S_\text{ALMA}/S_\text{ Planck }$} & \colhead{Notes}\\
	}
	\decimals
	\startdata
	1&Eq. \ref{eq:ALMA_v_Planck_slope_raw}	& 0.985 &$\pm$ &0.033 &$\pm$ &0.016 & Multiple step ``bridge" (Planck to ACT to ALMA) to minimize variability \\
	&	& (1.001 &$\pm$ & 0.035 &$\pm$ &0.015) & \\
	2&see \ref{subsubsec:act_bridge2ALMA} & 0.906 &$\pm$& 0.033 &$\pm$ &0.016 &  Omit ACTall from bridge.\\
	3&see \ref{subsubsec:act_bridge2ALMA} & 1.008 &$\pm$& 0.028 &$\pm$ &0.012 &  Omit MBAC from bridge.\\
	4&see \ref{subsubsec:act_bridge2ALMA} & 0.952 &$\pm$& 0.039 &$\pm$ &0.014 &  Omit season 16 PA2 from bridge.\\
	5&see \ref{subsubsec:act_bridge2ALMA} & 0.973 &$\pm$& 0.036 &$\pm$ &0.010 &  Omit both season 16 PA2 and MBAC from bridge.\\
	&     & (0.968 &$\pm$ & 0.034 &$\pm$ &0.012) & \\
	6&Eq. \ref{eq:ALMA_v_Planck_omit_MBAC_ACTall}  & 0.989 &$\pm$& 0.028 &$\pm$ &0.012   & Omit both ACTall and MBAC from bridge.\\
	&	& (1.004 &$\pm$& 0.029 &$\pm$ &0.010) & \\
	7&Eq. \ref{eq:ACT_v_Plack_slope_90GHz}  & 1.008 &$\pm$& 0.030 &$\pm$ &0.009 & Use 93 GHz ACT data, 100 GHz Planck data and ALMA spectral indices. \\
	&   & (1.026 &$\pm$& 0.030 &$\pm$ &0.010) & \\
	8&see \ref{subsubsec:act_alma_comp_w_act_alpha}  & 1.008 &$\pm$& 0.033 &$\pm$ &0.015 & Test ALMA 91.5 GHz calibration using ACT spectral indices.\\
	&	&  (1.027 &$\pm$& 0.034 &$\pm$ &0.016) &  \\
	9&see \ref{subsubsec:act_alma_comp_w_act_alpha}  & 1.008 &$\pm$& 0.034 &$\pm$ &0.013 & Test ALMA 103.5 GHz calibration using ACT spectral indices.\\
	&	& (1.021 &$\pm$& 0.035 &$\pm$ &0.013) &  \\
	10&see \ref{subsec:SPT_comp} & 0.968 &$\pm$& 0.032 &$\pm$ &0.006 & Use SPT 97.43 GHz not ACT as bridge.\\
	&   & (0.962 &$\pm$& 0.034 &$\pm$ &0.005) &  \\
	\enddata
\end{deluxetable*}

The most important result in this paper is the confirmation of the calibration accuracy of the ALMA flux density scale.  We employed a variety of means to link ALMA observations to the earlier, absolutely calibrated, Planck results, generally using ACT measurements as a bridge.  These results appear in Eqns. \ref{eq:ALMA_v_Planck_slope_raw}, \ref{eq:ALMA_v_Planck_omit_MBAC_ACTall}, \ref{eq:ACT_v_Plack_slope_90GHz} and \ref{eq:ALMAall_v_Planck} among others.  For convenience, we assemble in Table \ref{table:summary_results} the many results from the analysis in section \ref{sec:planck_based_calbration_ALMA}. The tests of ALMA calibration at 91.5 and 103.5 GHz are presented in finer detail here than in the text of section \ref{subsubsec:act_alma_comp_w_act_alpha}.  Although these comparisons of ALMA and Planck measurements were obtained from different data sets in many different combinations, the values are in general internally consistent with S(ALMA)/S(Planck) = 0.99, and in all but one case they are consistent with unity as well, thus confirming the absolute accuracy of ALMA in band 3.

\subsection{Combining results}\label{subsec:combining_results}

Readers may wish to weight or combine the results in different ways, but we suggest a reasonable summary of all the tabulated results is that ALMA Band 3 (84-116 GHz) calibration is consistent with Planck, with an uncertainty close to 2\%. \added{For one combination of the data, we find the ratio S(ALMA)/S(Planck) = 0.996 and conservatively estimate the uncertainty to be $\pm0.024$ (see below for details on this ''best estimate")}.

If we simply take the unweighted average of all 17 entries in column 2 of Table \ref{table:summary_results}. we obtain 0.989.  Given the correlations between various data sets, we make no attempt to compute an uncertainty on this average.

The results based on the most careful comparison of data sets at close to the same epoch appear in line 1. Since the more inclusive ALMA measurements (``ALMAall") used in the fit shown in parentheses overlap with the dedicated ALMA runs, the two entries are not independent.  Consequently, we cannot merely average them.  Both, however, are consistent with unity and with our suggested value of 0.996.

As noted in section \ref{subsubsec:act_planck_comp}, for the results in line 1, we minimize the effects of variability by using several ACT data sets as a bridge from Planck to ALMA.  The uncertainty in each of these steps accumulates (though it is dominated by the uncertainty in the comparison of ACT MBAC data to Planck).  Lines 2, 3 and 4 in Table \ref{table:summary_results} represent attempts to reduce the number of steps (at the cost of larger and less controlled effects from variability of the sources employed). The unweighted average of these fits is 0.955.
Lines 5 and 6 take this approach further by using only a single ACT data set to link Planck to ALMA.  In both cases, we avoid using the small ACT MBAC survey, since it has only 33 sources in common with Planck.  By eliminating this step, we obtain many more matches between ACT and Planck; the larger number helps to average out the effect of variability and also to lower uncertainty.

Line 7 uses only 93 GHz ACT data, giving fits entirely independent of the others; both results are somewhat larger than unity, but consistent with it.  We may also form an average of all the fits that do NOT make use of the ACT 93 GHz data; that average is 0.976.
Lines 8 and 9 use only ACT 2016 data, and do not rely on Band 7 data from ALMA.  The results thus provide a check on those reported on line 1.  Line 10 lists results obtained using SPT measurements, rather than ACT, as the bridge between Planck and ALMA.
Finally, we consider just the three data sets with the least overlap, MBAC, ACTall, and season 16 ACT data at 93 GHz, and hence lines 5, 6, and 7 of Table \ref{table:summary_results}. The  unweighted average, including comparisons with both the dedicated ALMA observation and the larger ALMA data set, is 0.995.  If we take the inverse variance weighted average of these same values, we obtain 
\begin{equation}
    S(\text{ALMA})/S(\text{Planck}) = 0.996 \pm 0.018 \pm (0.006)
\end{equation}
\neww[The internal error cited here is based on the assumption that the three data sets are fully independent, but that the fits based on the dedicated ALMA measurements are fully degenerate with those based on ALMAall]{It should be noted that this does not represent the minimum variance estimate, given that we expect significant covariance between fits based on the dedicated ALMA measurements and the corresponding measurements employing ALMAall. The error was computed under the assumption that those measurements are fully degenerate, taking into account the resulting off-diagonal elements of the covariance matrix}.  The latter assumption is conservative, but still yields an error estimate lower than 0.03.

\subsection{Earlier results using Planck to calibrate ground-based instruments}\label{earlier}

An earlier paper \citep{louis_atacama_2014} compares the early Planck PCCS1 flux densities with preliminary MBAC measurements at 148 and 218 GHz. At both frequencies, the early ACT MBAC measurements ran 1-2\% high compared to Planck PCCS1 values after removing a few known variable sources from the fits. When we correct the Planck calibration to its final values using Tables \ref{table:calibration_table} and \ref{table:PCCS1_v_PCCS2_comp}, the results reported in 
 \citet{louis_atacama_2014} become:

\begin{eqnarray}
S(\text{MBAC})/S(\text{Plank 143 GHz})&=& 0.988\pm 0.021,\\ 
S(\text{MBAC})/S(\text{Planck 217 GHz})&=& 0.938\pm 0.031.\
\end{eqnarray}

The first of these is in acceptable agreement with the results of Eq. \ref{eq:MBAC_v_Plack_slope_final} here. As was the case for SPT (Eqn. \ref{eq:SPT215_v_PCCS2}), the ground-based ACT measurements at $\sim$217 GHz are lower than Planck values (by ${\sim}2\sigma$ for ACT).

\new[Finally,]{A comment on a related test of the accuracy of the calibration of the Jansky Very Large Array.} \citet{partridge_absolute_2016} earlier found that flux density measurements by the VLA ran a statistically significant ${\sim}6\%$ below Planck measurements at 43 GHz. As noted in section \ref{sec:calibraton_standards}, there have been small adjustments to Planck calibration since that earlier work.  Could these changes in Planck calibration explain or mitigate this discrepancy?  No; at 43 GHz, the changes in calibration and beam solid angle between the first and final Planck releases are at sub-percent levels and partially cancel. On the other hand, when the data from that work is treated with the methods used here, including ODR, the discrepancy is reduced to $4 \pm 1.3 \%$, smaller but still significant.  

\subsection{Other ALMA Bands}\label{subsec:alma_planck_direct}

We cannot employ ACT observations as a bridge to link Planck measurements at 353 GHz and ALMA observations at 343 GHz (Band 7) as we did at lower frequencies since ACTPol data are available only at 93 and ${\sim}145$ GHz.  Nevertheless, we can set some rough constraints on ALMA flux density calibration in this band.  Since we employed ALMA observations in Band 7  in section \ref{subsec:linking_planck_alma} to calculate spectral indices, we can claim that there is no gross discrepancy between ALMA Band 7 calibration and the absolute calibration of Planck. For instance, a 10\% error in calibration at 343.5 GHz would produce a $\sim$3\% mismatch in the ALMA-Planck calibration, or a roughly 1$\sigma$ shift in our results.

Since the absolute calibration of ALMA is based on observations of Uranus, our results also validate models of the microwave emission of Uranus \citep[e.g.][]{butler_flux_2012} near 100 GHz.  For other direct measurements of Uranus by WMAP and Planck see \cite{weiland_seven-year_2011} and \cite{planck_collaboration_planck_2017}, respectively.  

Unlike the case for lower frequencies, the Planck 545 GHz channel was also calibrated on Uranus (and Neptune), not the annual dipole  \citep{planck_collaboration_planck_2020}. In that paper, it is pointed out that the CMB dipole derived from this calibration at 545 GHz agrees with the dipole measured at lower frequencies to 1$\%$ precision.  This in turn implies that Uranus is well calibrated at 545 GHz, so the flux density scale based on Uranus is accurate in ALMA Bands 8 and 9 as well.

\section{Conclusions}\label{sec:conclusions}
Our goal in this paper is to test the accuracy of the calibration employed by ALMA by comparing ALMA measurements of extragalactic sources to measurements of those same sources by Planck, which is absolutely calibrated to sub-percent precision.  Since ALMA did not begin to observe until Planck had ceased operations, and most of these sources are variable, we introduced a number of procedures to minimize the impact of variability.  The most crucial step was the use of ACT and SPT observations as  “bridge” to link Planck results to the later ALMA results.

Although these comparisons of ALMA and Planck measurements were obtained from different data sets in many different combinations, as summarized in Table \ref{table:summary_results}, the values are in general internally consistent, and in all but one case they are consistent with unity as well, thus confirming the accuracy of ALMA calibration in Band 3 (84-116 GHz).  All the values listed in Table \ref{table:summary_results} lie well within the  $\pm5\%$ normally assumed for ALMA calibration uncertainty based on planet observations. 

In section \ref{subsec:combining_results}, we examine the consistency of our results.  In that same section, we take the inverse variance weighted average of 3 largely independent data sets to obtain our best estimate of the relative calibration: $S(\text{ALMA})/S(\text{Planck}) = 0.996 \pm 0.018 \pm (0.006)$.  The error cited is conservative, since it assumes complete degeneracy of overlapping data sets

We conclude that a ratio $S(\text{ALMA})/S(\text{Planck})  = 0.996 \pm 0.024$ is a reasonable summary of our findings. We suggest that $\pm2.4\%$ is a reasonable value to employ as the standard figure for ALMA flux density calibration uncertainty rather than the $\pm5\%$ now employed, at least in ALMA band 3. 

We also report no evidence for bias in the ALMA calibration at other frequencies.  Likewise, the calibration of both ACT and SPT at 90-150 GHz seems sound.  In contrast, we find a significant mismatch between ground-based ACT and SPT measurements at 220 GHz and the corresponding Planck values at 217 GHz.

\new[]{Finally, we note that new ACT surveys covering roughly half the sky are now available  \citep{naess_atacama_2020}.  These contain roughly 18,000 compact sources; catalogs of these sources and their measured flux densities will soon be available.  In addition, the ACT team is undertaking a comprehensive reassessment of ACT calibration, including more detailed calculations of beam solid angle.  These advances will soon allow a more precise comparison between ACT and ALMA.  The crucial link back to Planck, however, must still rely on the comparison between early MBAC observations and those made at the same time by Planck.} 

\section*{Acknowledgements}

Planck (\url{http://www.esa.int/Planck}) is a project of the European Space Agency (ESA) with instruments provided by two scientific consortia funded by ESA member states and led by Principal Investigators from France and Italy, telescope reflectors provided through a collaboration between ESA and a scientific consortium led and funded by Denmark, and additional contributions from NASA (USA).

ACT was supported by the U.S. National Science Foundation through awards AST-0408698, AST-0965625, and AST-1440226 for the ACT project, as well as awards PHY-0355328, PHY-0855887 and PHY-1214379. Funding was also provided by Princeton University, the University of Pennsylvania, and a Canada Foundation for Innovation (CFI) award to UBC. ACT operates in the Parque Astron\'omico Atacama in northern Chile under the auspices of the Comisi\'on Nacional de Investigaci\'on (CONICYT).  The development of multichroic detectors and lenses was supported by NASA grants NNX13AE56G and NNX14AB58G. Detector research at NIST was supported by the NIST Innovations in Measurement Science program.  We are deeply grateful to all members of the ACT team for the observations used here, and for advice and assistance as this paper was drafted.

We thank the ALMA calibration team for their efforts that contributed to this paper. This paper makes use of the following ALMA data: ADS/JAO.ALMA\#2011.0.00001.CAL. ALMA is a partnership of ESO (representing its member states), NSF (USA) and NINS (Japan), together with NRC (Canada), MOST and ASIAA (Taiwan), and KASI (Republic of Korea), in cooperation with the Republic of Chile. The Joint ALMA Observatory is operated by ESO, AUI/NRAO and NAOJ. The National Radio Astronomy Observatory is a facility of the National Science Foundation operated under cooperative agreement by Associated Universities, Inc.

GSF acknowledges support through the Isaac Newton Studentship and the Helen Stone Scholarship at the University of Cambridge.  BP thanks the Provost's Office at Haverford College for financial support.

In addition to the software cited in the body of the paper we acknowledge the use of  \textit{Python 3.7} and many of its standard packages \citep{van_rossum_python_1995,van_rossum_python_2009}. Furthermore, we use the scientific computing packages \textit{SciPy} \citep{virtanen_scipy_2020} and \textit{NumPy} \citep{travis_guide_2006,walt_numpy_2011} as well as \textit{Astropy} \citep{astropy_collaboration_astropy_2013, price-whelan_astropy_2018}.

\appendix

\section{Frequency-to-frequency consistency of Planck measurements of compact sources}\label{appendix:f2f_plack_consistency}

We provide here more detail on how we confirmed the channel-to-channel consistency of Planck flux density measurements on compact sources.  To test the consistency at 100, 143 and 217 GHz, for instance, we constructed a band-matched catalog of Planck sources from the PCCS2 catalog, again using a search radius of $6^\prime$.  We also corrected the PCCS2 flux densities by the very small factors shown in column 6 of Table \ref{table:calibration_table} to update them to the final Planck calibration.  We then color-corrected flux density measurements at 100 and 217 GHz for each source, and used the spectral index $\alpha$ derived from these corrected values to predict the flux density at the intermediate frequency of 143 GHz:
\begin{equation}
	S(143 \text{ GHz})_\text{pred} = S(100\text{ GHz})\left(\frac{143}{100}\right)^\alpha  
\end{equation}
These predicted values were then compared to the measured, color-corrected values at 143 GHz.  Note that this prediction does not take account of any curvature in the spectra of sources; we return to this point below.

Before making the comparison, we excluded as usual 33 bright or extended sources.  We also employed a filter on the spectral index to exclude sources with evidence of thermal emission in order to keep the color corrections small.  Specifically, we required the 100-143 GHz spectral indices to lie in the interval $-1 \leq \alpha < 0$.  That left us with 461 sources.

If the 3 Planck bands we are testing are consistent, we expect the measured and predicted flux densities to match on average (unit slope in the fit), absent spectral curvature.  In fact, we find $0.990 \pm 0.002$; the three Planck bands at 100, 143 and 217 GHz are consistent at the ${\sim}1$\% level.  The small amount of scatter is due in part to measurement error, but some may also be due to source variability (a given source swept through Planck’s beams for different frequencies at slightly different times).  Ten outliers were excluded by the iterative procedure described in section \ref{subsec:flux_comp_general_features}. 

To perform a similar check of the 217 GHz flux densities, we used color-corrected Planck data at 143 and 353 GHz to predict flux densities at 217 GHz, and compared these predictions to the observed flux densities.  There were fewer sources that met the spectral index selection criterion (148), and the observed slope was a bit above unity at $1.017 \pm 0.005$.   

Finally, we repeated this exercise at 100 GHz, starting with color-corrected flux densities at 70.4 and 143 GHz.  In this case, the measured 100 GHz flux densities were on average $2.7 \pm 0.3\%$ higher than the predicted values.  This result could be due to a slight curvature in the average SED of sources around 100 GHz, or possibly to contamination by Galactic CO ($1\to0$) line emission leaking into the Planck 100 GHz channel. We tested for the latter by excluding all sources with Galactic latitude below 20 degrees, on the assumption that Galactic CO contamination would be more prevalent at low latitudes \cite{planck_collaboration_planck_2020-1}.   The result was largely unchanged: the measured values remained higher by  $2.5 \pm  0.4$\%.   We thus conclude that spectral curvature is more likely the cause of the excess of measured over predicted flux.  If we assume a simple model based on a single, sharp break in the spectrum at 100 GHz, a small, average change of $\Delta\alpha = - 0.13$ at 100 GHz would explain the 2.7\% excess in measured flux densities we detect.  \added{A change of slope of this sign and magnitude is consistent with the steepening of spectral indices seen for a wider range of Planck sources \citep[e.g.][]{planck_collaboration_planck_2011-2}.}

We conclude that the catalogued Planck flux densities at 143,  217 and 353 GHz are consistent to 2\% or better, and that there is only a mild inconsistency at 100 GHz.

\section{Consistency of ACT Data across Arrays and Epochs}\label{appendix:act_updates_note}

As noted in section \ref{subsec:act_internal_consistence}, given the changes in the detectors employed at ACT (and some changes in the analysis pipeline as well), we sought to confirm that ACT flux density measurements of compact sources, once properly calibrated, were internally consistent across detector arrays, sky area and time.  For instance, observations during the 2014 season were made with two different, polarized receiver arrays (called PA1 and PA2), operating with center frequencies of 144.3 and 144.7 GHz, respectively \citep[for sources with spectral indices near $-0.5$; see ][]{choi_atacama_2020}.  For these comparisons, we arbitrarily selected PA2 as our standard for comparison. 

\begin{deluxetable}{cccDcDc}
	\tablecaption{Array-to-array consistency of ACT $\sim$145 GHz measurements of compact sources in a given season. Note the outlier, PA3 in 2016; we exclude it from all analyses here. \added{In the last column, we give the total number of sources used, with the number of sources removed during iteration in parentheses}.\label{table:ACT_array2array_consistency}}
	\tablehead{
		\colhead{Season} & \multicolumn2c{Arrays Compared} & \multicolumn5c{Comparison of  Overall }&\colhead{N}\\
		\colhead{}&\colhead{A}&\colhead{B}&\multicolumn5c{Calibration, $S(A)/S(B)$}&\colhead{}
	}
	\decimals
	\startdata
	2014 &PA1&PA2&$1.011$ & $\pm$ & $0.0018$ & 321(6) \\
	2015&PA1&PA2&$1.018$ & $\pm$ & $0.0013$ & 2319(70) \\
	2015&PA3&PA2&$1.019$ & $\pm$ & $0.0015$ & 2240(78) \\
	2015&PA3&PA1&$1.009$ & $\pm$ & $0.0010$ & 4517(208) \\
	2016&PA3&PA2&$1.073$ & $\pm$ & $0.0018$ & 3371(59)
	\enddata
\end{deluxetable}

In 2014, two disjoint sky regions were observed; we combined source lists, and as usual dropped 33 bright or extended sources.  For all comparisons, we also excluded numerous weak sources by fixing the threshold to 20 mJy.  That left 323 sources in all for 2014: when we compared observations made by the two arrays, the slope was close to unity at $1.011 \pm 0.0018$. There is, as expected, much less scatter than in comparisons across time, since source variability is strongly reduced.  (Variability could enter at a low level: while the measurements by the two arrays were made during the same season, they were not exactly simultaneous because of different cuts to the data.)  The remaining scatter results from measurement error (small in this case since we used only sources with $S > 20$ mJy, or $>10\sigma$).  We conclude that results from the two detector arrays used in 2014 are consistent at the $\sim$1\% level. In 2015, the agreement was not as close; we found $S(\rm{PA1})/S(\rm{PA2}) = 1.018 \pm 0.0013$.  

PA3, added in 2015, showed a greater difference at $\sim$145 GHz.  We again restricted the comparison to sources with $S > 20$ mJy.  In 2015, where this threshold was $\sim$10 times the noise, the fit was acceptable at $1.019 \pm 0.0015$ and $1.009 \pm 0.0010$ when compared to PA2 or PA1, respectively.  In 2016, however, the survey was much wider in area and shallower, so 20 mJy was closer to the noise level.  Here PA3 was sharply discrepant.  Note that we did test PA3 against PA2 in 2016 with a higher threshold of 50 mJy, but a substantial discrepancy remained. Hence we excluded 2016 PA3 145 GHz data from all further considerations.  

\begin{deluxetable}{llDcDc}[t]
	\tablecaption{Season-to-season consistency of ACT measurements of compact sources at $\sim$145 GHz (93 GHz on last line).  Recall that we do not use PA3 results at 145 GHz from 2016. \new[]{In 2015, there were few and scattered 93 GHz observations, so we rely on the 2016 measurements.  \added{As earlier, we indicate the total number of sources used with the number of sources removed during iteration ("outliers'') in parentheses.  The number of outliers is omitted in line 2 since we made this comparison without iteration given that the data sets partially overlap, i.e. observations for PA2 2016 are contained within ``ACTall".}} \label{table:ACT_season2season_consistency}}
	\tablehead{
		\multicolumn2c{Array(s)/Seasons(s) compared} & \multicolumn5c{Comparison of  Overall }&\colhead{N}\\
		\colhead{A}&\colhead{B}&\multicolumn5c{Calibration, $S(A)/S(B)$}&\colhead{}
	}
	\decimals
	\startdata
	MBAC (2008-10) & ACTall (2013-16) & $1.0066$ & $\pm$ & $0.0116$ & 284(30) \\
	PA2 2016 & ACTall (2013-16) & $1.0102$ & $\pm$ & $0.0030$ & 3132\\
	PA1 2013 & PA1 2015 & $1.060$ & $\pm$ & $0.015$ & 173(23)\\
	PA1 2014 & PA1 2015 & $1.028$ & $\pm$ & $0.008$ & 306(42)\\
	PA2 2014 & PA2 2015 & $1.039$ & $\pm$ & $0.008$ & 283(34)\\
	PA2 2016 & PA2 2015 & $0.995$ & $\pm$ & $0.005$ & 1260(112)\\
	PA3 2016 & PA3 2015 & $1.098$ & $\pm$ & $0.010$ & 1161(64)\\
	93 GHz 2015 & 93 GHz 2016 & $0.966$ & $\pm$ & $0.006$ & 1098(109)\\
	\enddata
\end{deluxetable}

More important for the use of ACT observations as a “bridge” to connect Planck and ALMA is their consistency from season to season.  Table \ref{table:ACT_season2season_consistency} displays the results of comparing measurements of compact sources with flux density $>20$ mJy for various arrays across different ACT observing seasons.  Here, of course, variability introduced more scatter and hence larger uncertainties than in Table \ref{table:ACT_array2array_consistency}.  Also shown are comparisons between some individual seasons and the weighted 2013—2016 average, ACTall, (taken from section \ref{subsec:act_internal_consistence}).  Note that PA 3 at 144.1 GHz is again discrepant. At 93 GHz, the agreement between 2015 and 2016 PA3 observations was better, but still not consistent with unity.  The ${\sim} 4 \sigma$ discrepancy between 2013 and 2015 PA1 measurements can be ascribed to both the longer interval between observations and the small number of sources in the 2013 observations. 

\section{Variability bias}\label{appendix:var_bias}

In this appendix we consider the potential bias introduced by source variability in the comparison of flux densities from two catalogs with very different thresholds. That is the case for Planck and ACT; the catalog thresholds differ by a factor of ${\sim}20$.  Consequently, a variable source that happened to be bright when Planck observed it, but faded for the later ACT observations, would still be detected by ACT given its much greater sensitivity.  But the opposite is not true.  Hence a simple list of matching sources could be biased.  One remedy, to use a third, “neutral” catalog to select sources, is not available at the frequencies we consider; the highest frequency wide-area radio survey is at 20 GHz.  We consider this potential bias in general terms, then describe the specific steps we took to minimize in the present work.

 \subsection{Using cuts in flux density as a remedy.}

Consider two catalogs, one made at epoch $t_1$, with noise level $N_1$, and a nominal 100\% completeness at $S_1$, and the second catalog with $t_2$, $N_2$, and $S_2$.  Also, let $N_2$ be $<< N_1$, as is the case for ACT and Planck.  In principle, the bias we are considering could be removed by setting a threshold on the  flux densities from catalog 2 so high that every remaining source is matched in the noisier catalog 1.  In the context of this paper, however, that would mean dropping nearly half of the 33 entries in Table \ref{table:PCCS1_v_MBAC_comp}.

An alternative is to fix a threshold based on the nominal completeness limit of catalog 1, dropping all sources from catalog 2 with $S < S_1$.  In the absence of variability, this would produce an unbiased set of matches.  It is possible, however, for a source to have flux density $S > S_1$ at epoch $t_2$, but a much lower flux density at epoch $t_1$, and hence to be absent in catalog 1.  We call these ``missing sources."  As already noted, simply ignoring these would bias the comparison.   Can we place any limits on the flux density of these missing sources at $t_1$?   The flux densities of the missing sources at $t_1$ cannot in principle exceed $S_1$, providing an upper limit.  On the other hand, in principle there is no lower limit, but we know it is unusual for sources to fade by more than a factor 10.  If we knew the probability of fractional variation P(V) for the class of sources we are considering, we could establish a reasonable lower limit as follows. Say that 90\% of variable sources have $V < V_t$,  then adopt (1-$V_t$) times the observed flux density in catalog 2 as the probable lower limit on the flux of  each source missing from catalog 1 (and $S_1$ as the upper limit).  If there are many missing sources, lower limits could be assigned by drawing from P(V).  A further refinement would be to consider the completeness curve for catalog 1.

In this paper, since the number of matched  sources is relatively small, we adopt a simpler approach.   To retain more sources when we compare Planck and ACT flux densities, we apply a threshold to catalog 2 (the ACT catalog) corresponding not to the 100\% completeness level of catalog 1 (Planck) but instead to the 90\% completeness level.  Hence we cut the ACT  catalog at 220 mJy rather than 320 mJy.  We find that only 3 Planck sources are missing in the sense just defined.  Since only 3 of 33 are missing, we do not bother with a full treatment using P(V).  Instead we fix one limit by assigning each of the 3 missing sources the maximum flux density $S_1$, in this case 320 mJy, and assign a typical Planck uncertainty of $\pm40$ mJy to it. By assigning the maximum flux density to each missing Planck source, we establish a nominal lower limit on the slope of ACT versus Planck comparisons, like those shown in Fig.3. The resulting slope is $1.0251 \pm 0.0258 \pm (0.0150)$.

To fix the other limit, we make use of the finding in \cite{datta_atacama_2019} that only ${\sim}25$\% of  ACT sources at 144 GHz vary by more than 40\% over a five year time span. We then set the Planck flux density for all 3 missing sources to (1-V) = 0.6 of  the corresponding ACT flux density. For computational ease, we use the same value for each of the three missing sources: 190 mJy.  With lower values inserted for the flux of the missing Planck sources, the slope is $1.0176 \pm 0.0308 \pm (0.0140)$. The true slope likely lies between these two values. \added{While the putative lower limit on the slope is a bit higher than the upper limit, the difference is well within the margin of error.} We therefore adopt an intermediate, inverse-variance weighted, value of $1.0220 \pm 0.0283 \pm (0.0148)$ as our standard value. 

We took these steps since simply omitting the sources with missing Planck flux densities might have biased the result. To check this, we re-ran the ACT- Planck comparison without the three missing Planck sources: as expected, the slope was slightly lower, but still consistent, at $1.0079 \pm 0.0301 \pm (0.0167)$.

Given the central role of this ACT-Planck comparison, we also made a number of other tests. First, as usual, we cut the ACT catalog at 220 mJy, but replaced the 3 missing source with an intermediate flux density of $220 \pm 40$ mJy, changing the slope to $1.0149 \pm 0.2981 \pm (0.0161)$. We also tried cutting the ACT catalog at a lower flux (190 mJy) in order to include a few more sources (39 rather than 33 matches). In this case, 6 of 39 Planck sources were missing. Replacing the Planck flux density of missing sources with 320 mJy yielded a (lower limit) slope of $0.9945 \pm 0.0277 \pm (0.0156)$; using (1-V) = 0.6 instead yielded $1.0325 \pm 0.0262 \pm (0.0132)$.  Note that we gained only a slight improvement in uncertainty by including more sources.  \added{Finally, we compared flux densities for a smaller number of bright sources, 19 sources with flux density above 320 mJy.  This lowered the slope by less than $1\sigma$ to 0.983, but increased the statistical error to $\pm 0.0428$.}

\bibliography{ACT-ALMA-PLanck_references}
\bibliographystyle{aasjournal}


\listofchanges

\end{document}